\documentclass[%
 superscriptaddress,
 elsarticle,
 amsmath,amssymb,
reprint,%
 author-year,%
longbibliography
]{revtex4-1}

\usepackage{graphicx,dcolumn,bm}
\usepackage{hyperref}
\usepackage{graphicx,dcolumn,amsmath,amssymb,color,mathrsfs,titlesec,hyperref,enumerate,bm,subfigure,float,comment}
\definecolor{linkColor}{rgb}{1,0,0}
\hypersetup{pdfborder={0 0 0},colorlinks=true,urlcolor=linkColor,citecolor=linkColor}

\usepackage{xargs} 
\newcommand{\vect}[1]{\mathbf{#1}}
\newcommand{\ii}{{i\mkern1mu}}

\emergencystretch=\maxdimen
\begin{document}
\title[py4DSTEM]{py4DSTEM: a software package for multimodal analysis of four-dimensional scanning transmission electron microscopy datasets}

\author{Benjamin H Savitzky}
\email{ben.savitzky@gmail.com}
\affiliation{National Center for Electron Microscopy, Molecular Foundry, Lawrence Berkeley National Laboratory, 1 Cyclotron Road, Berkeley, CA, USA, 94720}

\author{Lauren A Hughes}
\affiliation{National Center for Electron Microscopy, Molecular Foundry, Lawrence Berkeley National Laboratory, 1 Cyclotron Road, Berkeley, CA, USA, 94720}

\author{Steven E Zeltmann}
\affiliation{National Center for Electron Microscopy, Molecular Foundry, Lawrence Berkeley National Laboratory, 1 Cyclotron Road, Berkeley, CA, USA, 94720}

\author{Hamish G Brown}
\affiliation{National Center for Electron Microscopy, Molecular Foundry, Lawrence Berkeley National Laboratory, 1 Cyclotron Road, Berkeley, CA, USA, 94720}

\author{Shiteng Zhao}
\affiliation{National Center for Electron Microscopy, Molecular Foundry, Lawrence Berkeley National Laboratory, 1 Cyclotron Road, Berkeley, CA, USA, 94720}

\author{Philipp M Pelz}
\affiliation{National Center for Electron Microscopy, Molecular Foundry, Lawrence Berkeley National Laboratory, 1 Cyclotron Road, Berkeley, CA, USA, 94720}
\affiliation{Department of Materials Science and Engineering, University of California, Berkeley, Berkeley, USA, 94720}

\author{Edward S Barnard}
\affiliation{Molecular Foundry, Lawrence Berkeley National Laboratory, 1 Cyclotron Road, Berkeley, CA, USA, 94720}

\author{Jennifer Donohue}
\affiliation{National Center for Electron Microscopy, Molecular Foundry, Lawrence Berkeley National Laboratory, 1 Cyclotron Road, Berkeley, CA, USA, 94720}

\author{Luis Rangel DaCosta}
\affiliation{Department of Materials Science and Engineering, University of Michigan, Ann Arbor, Michigan, USA, 48109}

\author{Thomas C. Pekin}
\affiliation{Institut f{\"u}r Physik, Humboldt-Universit{\"a}t zu Berlin, Newtonstra{\ss}e 15, 12489 Berlin, Germany}

\author{Ellis Kennedy}
\affiliation{National Center for Electron Microscopy, Molecular Foundry, Lawrence Berkeley National Laboratory, 1 Cyclotron Road, Berkeley, CA, USA, 94720}

\author{Matthew T Janish}
\affiliation{Los Alamos National Laboratory, Los Alamos, NM, USA, 87545}

\author{Matthew M Schneider}
\affiliation{Los Alamos National Laboratory, Los Alamos, NM, USA, 87545}

\author{Patrick Herring}
\affiliation{Toyota Research Institute, Los Altos, CA, USA, 94022}

\author{Chirranjeevi Gopal}
\affiliation{Toyota Research Institute, Los Altos, CA, USA, 94022}

\author{Abraham Anapolsky}
\affiliation{Toyota Research Institute, Los Altos, CA, USA, 94022}

\author{Peter Ercius}
\affiliation{National Center for Electron Microscopy, Molecular Foundry, Lawrence Berkeley National Laboratory, 1 Cyclotron Road, Berkeley, CA, USA, 94720}

\author{Mary Scott}
\affiliation{National Center for Electron Microscopy, Molecular Foundry, Lawrence Berkeley National Laboratory, 1 Cyclotron Road, Berkeley, CA, USA, 94720}
\affiliation{Department of Materials Science and Engineering, University of California, Berkeley, Berkeley, USA, 94720}

\author{Jim Ciston}
\affiliation{National Center for Electron Microscopy, Molecular Foundry, Lawrence Berkeley National Laboratory, 1 Cyclotron Road, Berkeley, CA, USA, 94720}

\author{Andrew M Minor}
\affiliation{National Center for Electron Microscopy, Molecular Foundry, Lawrence Berkeley National Laboratory, 1 Cyclotron Road, Berkeley, CA, USA, 94720}
\affiliation{Department of Materials Science and Engineering, University of California, Berkeley, Berkeley, USA, 94720}

\author{Colin Ophus}
\email{cophus@gmail.com}
\affiliation{National Center for Electron Microscopy, Molecular Foundry, Lawrence Berkeley National Laboratory, 1 Cyclotron Road, Berkeley, CA, USA, 94720}

\date{\today}
\begin{abstract}
Scanning transmission electron microscopy (STEM) allows for imaging, diffraction, and spectroscopy of materials on length scales ranging from microns to atoms.
By using a high-speed, direct electron detector, it is now possible to record a full 2D image of the diffracted electron beam at each probe position, typically a 2D grid of probe positions.
These 4D-STEM datasets are rich in information, including signatures of the local structure, orientation, deformation, electromagnetic fields and other sample-dependent properties.
However, extracting this information requires complex analysis pipelines, from data wrangling to calibration to analysis to visualization, all while maintaining robustness against imaging distortions and artifacts.
In this paper, we present py4DSTEM, an analysis toolkit for measuring material properties from 4D-STEM datasets, written in the Python language and released with an open source license.
We describe the algorithmic steps for dataset calibration and various 4D-STEM property measurements in detail, and present results from several experimental datasets.
We have also implemented a simple and universal file format appropriate for electron microscopy data in py4DSTEM, which uses the open source HDF5 standard.
We hope this tool will benefit the research community, helps to move the developing standards for data and computational methods in electron microscopy, and invite the community to contribute to this ongoing, fully open-source project.




\end{abstract}
\pacs{PACS Numbers}
\maketitle


\section{Introduction}
\label{S:introduction}

\begin{figure*}[htbp]
    \centering
        \includegraphics[width=\textwidth]{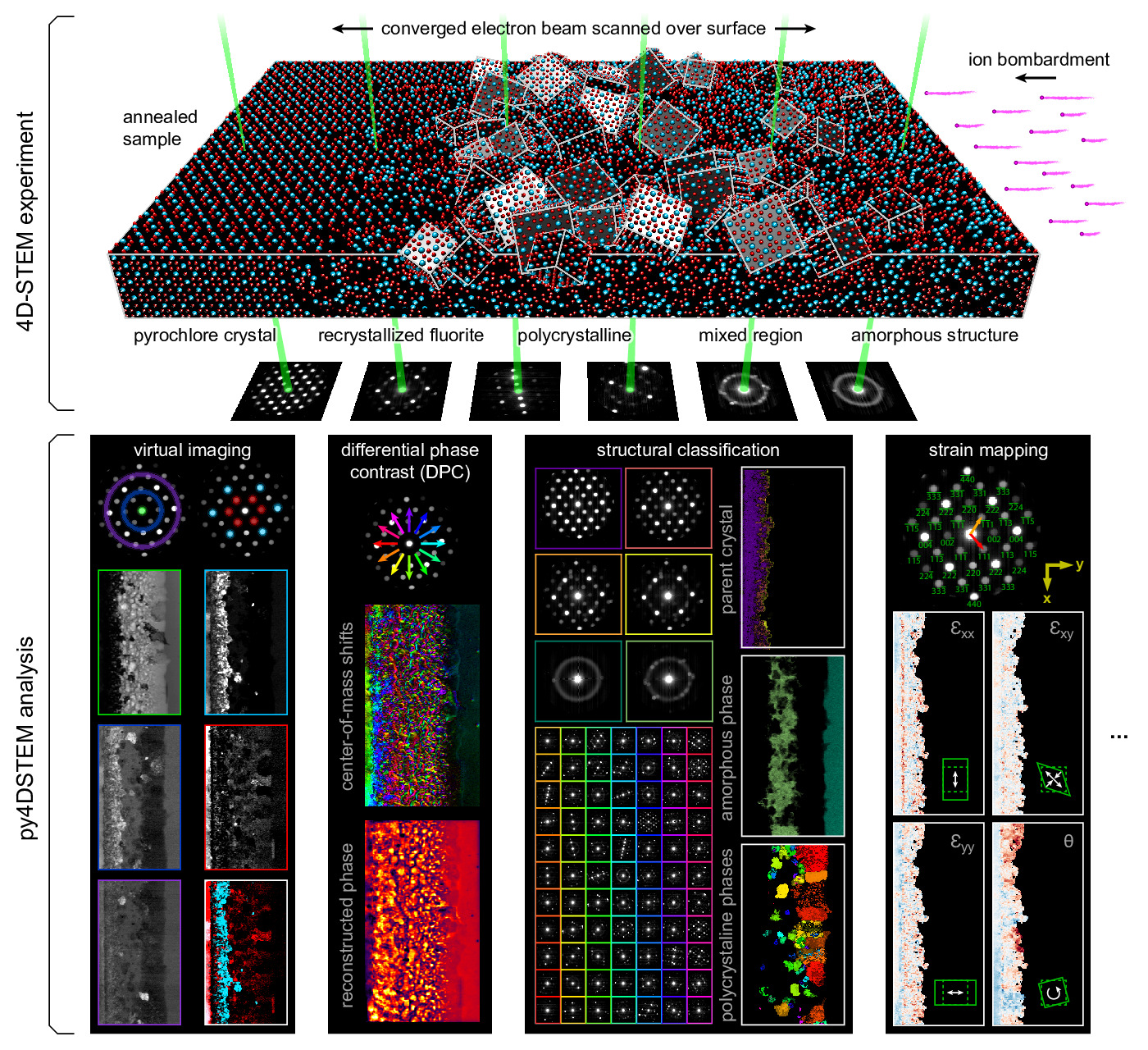}
    \caption{4D-STEM experimental geometry, and multimodal data analysis with py4DSTEM. An irradiated Gd$_2$Ti$_2$O$_7$ sample contains complex, nanoscale structure, apparent in the distinct electron diffraction patterns across the field-of-view.  From a single 4D-STEM experiment, py4DSTEM enables a range of measurements to be performed in post-processing, including virtual imaging, differential phase contrast, structural classification, strain mapping, and much more.}
	\label{F:superfigure}
\end{figure*}

In a scanning transmission electron microscopy (STEM) experiment, a beam of high energy electrons is focused to a very fine probe -- on the order of or, often, smaller than the atomic lattice spacing -- and rastered across the surface of the sample \cite{pennycook2011_STEM_textbook}.
In traditional STEM, a (two-dimensional) image is formed by populating the value of each pixel with the number of electrons (times a scaling factor) scattered into a detector at each beam position.
The geometry of the detector -- its size, shape, and position in the microscope's diffraction plane -- determines which electrons are collected at each probe position.
As a result, different detector geometries can give rise to rather different images, by varying which electron scattering processes dominate image contrast \cite{cowley1976scanning}.
A point detector placed on the optic axis yields a bright-field STEM image which is formally equivalent by reciprocity to a TEM image.
In contrast, annular detectors with large inner-radii are dominated by high momentum-transfer elastic scattering events, making high-angle annular dark-field STEM a popular geometry because image contrast then generally scales monotonically with the projected potential of the sample (``Z-contrast'' imaging) \cite{wang1989simulating}.
Low-angle annular detectors have greater sensitivity to lighter elements, but lose the advantage of simple Z-contrast interpretability due to the increased importance of phase contrast, i.e. self-interference of the electron beam wavefunction.
Many more detector geometries are possible, each best suited to reveal different aspects of sample structure, each suffering from different limitations \cite{spurgeon2020scanning}.

In four dimensional STEM (4D-STEM), we replace the standard STEM detectors, which integrate all electrons scattered over a large region, with a pixelated detector that captures the electron flux scattered to each angle in the diffraction plane \cite{zaluzec2002quantitative, fundenberger2003polycrystal, watanabe2007development, lupini2015ptychographic, tate2016high, ophus2019_4DSTEM_review}.
While a typical STEM image therefore produces a single number for each position of the electron beam, a 4D-STEM dataset produces a two dimensional image of diffraction space intensities for each real space beam position.
The resulting four-dimensional data hypercube can be collapsed in real space to yield information comparable to more traditional electron diffraction experiments.
Alternatively, it can be collapsed in diffraction space to yield a variety of ``virtual images,'' corresponding to both traditional STEM imaging modes as well as more exotic virtual imaging modalities \cite{schaffer2008hyperspectral, tao2009_SEND, gammer2015diffraction, zhang2017electroplating, hachtel2018sub}.
More information still can be extracted by judicious combination of real and reciprocal space.
The structure, symmetries, and spacings of Bragg disks can be used to extract spatially resolved maps of crystallinity, grain orientations, and lattice strain \cite{schwarzer1998automated, usuda2005strain, beche2009improved, caswell2009high, koch2012quantitative, kobler2013combination, pekin2017strain, hou2019metal}.
Redundant information in overlapping Bragg disks can be leveraged to deconvolve the electron beam shape from the sample structure, yielding the sample potential itself \cite{lazic2016phase, chen2016practical, muller2018atomic}.
Rings of diffracted intensity characteristic of amorphous samples can be used to extract correlation functions describing the short and medium range order and disorder.
Indeed, the space of possible quantities of physical interest which can be extracted from a single 4D-STEM experiment is formidable, leading others to use the term ``universal detectors'' for 4D-STEM capable pixelated cameras \cite{hachtel2018sub}. Fig.~\ref{F:superfigure} shows the experimental geometry of a 4D-STEM experiment, and various measurements performed from the same experimental dataset.
For a mathematical discussion of STEM and 4D-STEM image formation, see Appendix~\ref{A:STEM}.

The price paid for the versatility of 4D-STEM is new complexity in both the raw experimental data and in the computational processing required to extract meaningful measurements.
Maximizing the impact this new generation of STEM experiments will have on structural characterization research now requires that the computer processing methods which enable the various 4D-STEM characterization modalities are accessible to as broad and diverse a segment of the scientific community as possible.
Fortunately, a new generation of open source tools for electron scattering experiments is presently on the rise, such as hyperspy, pyXem, liberTEM, ncempy, and others\cite{pena2020_hyperspy,nord2019fast,johnstone2019pyxem,clausen2020_libertem}.


Here, we present free and open source software for analysis of 4D-STEM data.
The aim of the Python-based project py4DSTEM is threefold:
(1) to make 4D-STEM data analysis easy and accessible for everyone;
(2) to facilitate reproducibility, even in cases of complicated or multi-step processing workflows; and
(3) to provide a comprehensive, robust suite of 4D-STEM analysis tools, enabling high throughput, multimodal analysis in which a single dataset can simultaneously provide many distinct measurements of sample structure.
For ease and accessibility, py4DSTEM includes a complete API with associated documentation pages, many fully worked examples in the form of fully commented and interactive Jupyter notebooks, and a graphical user interface for fast data visualization and interaction.
For reproducibility, py4DSTEM defines a set of structured data object types for 4D-STEM data processing, establishes a set of HDF5-based file format conventions for 4D-STEM data, and makes it easy to release, with any publication, the complete and fully transparent code which generates results and figures from raw data.
For multimodal, high-throughput analysis, py4DSTEM includes a comprehensive suite of tools for structural analysis in crystalline and amorphous materials, including virtual imaging, phase and orientation mapping, strain mapping, radial distribution analysis, phase contrast imaging, classification, and more.
A self-consistent framework allows many or even all of these measurements to be readily performed on a single dataset.
The API and sample code for various analysis pipelines are freely available from the  \href{https://github.com/py4dstem/py4DSTEM}{py4DSTEM repository} \cite{savitzky2020py4dstem}.

The organization of this document is as follows:
following this introduction, Sec.~\ref{S:4D-STEM_data} discusses the nature of 4D-STEM data, and how data is structured in py4DSTEM.
Section~\ref{S:basic_processing} discusses basic processing algorithms which will typically be performed as precursors to the final measurements of interest, including locating Bragg disks, calibration, polar transformations, and classification.
Section~\ref{S:measurements_and_applications} covers various 4D-STEM measurements that can be performed in py4DSTEM, including virtual imaging, phase mapping, strain mapping in amorphous or crystalline materials, short and medium range order analysis in amorphous materials, and phase retrieval in very thin samples.
Conclusions are in Sec.~\ref{S:conclusions}.
Throughout, we have aimed to keep discussion qualitative in the main text, and have also included mathematical details for the interested reader in a number of appendices, referenced in the relevant sections.


\section{4D-STEM Data}
\label{S:4D-STEM_data}

Fundamentally, most 4D-STEM is just many electron diffraction experiments being run sequentially.
The nature of the diffraction pattern obtained at each scan position depends on the sample structure and the illumination conditions of the microscope, as illustrated schematically in Fig.~\ref{F:superfigure}.
In crystalline materials and with small-angle illumination, the periodic structure of the sample gives rise to a periodic pattern of disks in the diffraction plane \cite{carter2016transmission}.
A bright disk appears wherever the Bragg condition is met, with the disk positions reflecting a slice through the reciprocal lattice of the crystal.
In amorphous materials, concentric rings of diffuse intensity appear centered about the optic axis \cite{egami2003underneath}.
The radii of these rings reflect the characteristic spacings of the atoms in the sample, and can therefore be used to extract statistical measures of structure, such as the radial distribution function.
In analyzing crystalline materials, the crux of the analysis will generally be measuring the Bragg angles in each diffraction pattern, by determining the positions of all the Bragg disks.
In analyzing amorphous materials, analysis will generally revolve around radial integration of the diffraction patterns.
In samples containing both crystalline and amorphous regions, both types of analysis can be performed in concert.

\subsection{Experimental Conditions}

A complete discussion of the many experimental conditions to attend to in devising a given 4D-STEM experiment is beyond our scope, however, there is one parameter which stands apart in its centrality to both acquiring and understanding 4D-STEM data: the convergence semiangle, $\alpha$.
When examining a diffraction pattern, $\alpha$ corresponds to the radius of the bright-field disk in the diffraction plane, and therefore also the radius of each refracted Bragg disk in a crystalline sample.
In real space, the probe size is inversely related to $\alpha$; larger convergence angles correspond to finer probes, and overlapping disks are required to generate sub-lattice sized probes and therefore allow atomic resolution imaging \cite{kirkland2010advanced}.
In extracting a strain map, for example, non-overlapping disks are important, both to facilitate the detection of the disk positions, and also because strain is a physical quantity only defined on length scales equal to or larger than single unit cells.
For a ptychographic reconstruction of the atomic potentials of very thin materials, overlapped disks are essential, as they provide the redundant information required to extract the phase of the electron wavefunction and the sample electrostatic potential \cite{hegerl1970dynamische}.  
For analysis of amorphous materials, measuring radial distribution functions requires nearly-parallel illumination (a small semiconvergence angle), while measurements of medium range order in fluctuation electron microscopy experiments will often vary the probe semiangle to probe different sizes of atomic clusters \cite{rodenburg1999measurement, mu2016radial}.
In general, the convergence angle should be selected carefully in light of the particular requirements of the experiment.



\subsection{Multimodal analysis - one dataset, many measurements}
\label{S:multimodal_analysis}

A major advantage of 4D-STEM is the ability to perform a single experiment from which many distinctly meaningful structural measurements can be made.
We take as our guiding example the Gd$_2$Ti$_2$O$_7$ (GTO) crystal shown in Fig.~\ref{F:superfigure}.
A pyrochlore structured GTO single crystal was first bombarded with ions, creating an amorphized layer.
Then the sample was annealed, creating both a layer of recrystallization on the parent lattice as well as a band of smaller crystallites embedded in an amorphous matrix.
Each of these regions is clearly visible in the diffraction patterns associated with various beam positions of the 4D-STEM scan.

A selection of the types of measurements that can be performed from this dataset are shown in the figure.
These include:
virtual imaging spanning bright field images, annular dark field images, and dark field images of individual or multiple Bragg reflections (see Sec.~\ref{S:virtual_imaging});
differential phase contrast imaging, whereby shifts in the center of mass of the beam are used to back out the sample structure\footnote{Note that while DPC provides useful image contrast in a fairly wide array of contexts, physical interpretation, and in particular interpretation in terms of the local sample potential, should be undertaken with care.  In this dataset, for instance, the presence of non-overlapping Bragg disks indicates that there exists sample structure (the atomic lattice) which is too fine for our probe to resolve, and which will not be reflected in a DPC reconstruction.  Moreover, the spatial sampling here is larger than probe width, so any variation in the potential between sampling points will be effaced in the reconstruction.  Thus this DPC image, though still informative to a point, has no simple physical interpretation.  Images in this category might be referred to as ``pseudo-DPC''.  For more discussion, and an example where the DPC image is well thought of as a reconstructed sample potential, see Sec.~\ref{S:DPC} and Appendix~\ref{A:dpc}.} (see Sec.~\ref{S:DPC});
strain mapping, showing the local deformations of the atomic lattice (see Sec.~\ref{S:crystalline_strain_mapping});
and structural classification, where regions of distinct structure are identified and segmented (see Secs.~\ref{S:classification} and \ref{S:phase_mapping}).
With py4DSTEM, these analyses and more can all be applied to a dataset within a single, unified framework.

\subsection{Data Structures}
\label{S:datastructure}

\begin{figure}[htbp]
    \centering
    \includegraphics[width=\columnwidth]{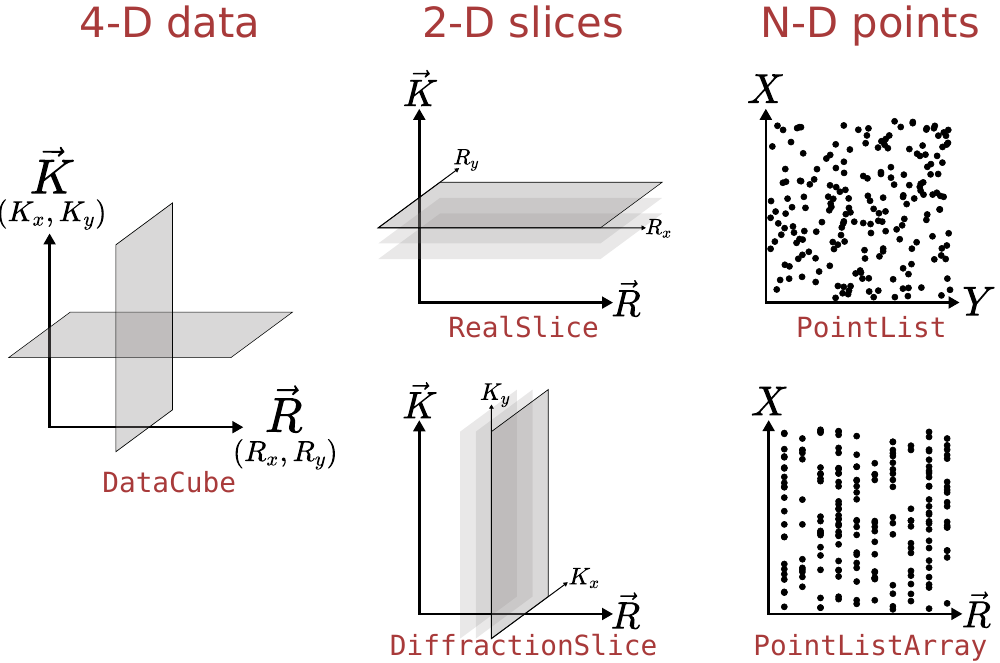}
    \caption{py4DSTEM data structures.  Data is saved as one of 5 classes of dataobjects -- \texttt{DataCube}, \texttt{DiffractionSlice}, \texttt{RealSlice}, \texttt{PointList}, and \texttt{PointListArray} objects.}
    \label{F:datastructures}
\end{figure}

Data in py4DSTEM is structured in five different types, broadly distinguished by their dimensionality, shown in Fig.~\ref{F:datastructures}.
In-program, these are implemented as the following Python classes: \texttt{DataCube}, \texttt{DiffractionSlice}, \texttt{RealSlice}, \texttt{PointList}, and \texttt{PointListArray}.
\texttt{DataCube} instances contain a 4D data array corresponding to the complete 4D-STEM dataset.
\texttt{DiffractionSlice} and \texttt{RealSlice} instances contain one or more 2D arrays with shapes corresponding to that of diffraction space (i.e. the detector shape) or of real space (i.e. the raster scan shape), respectively.
A \texttt{DiffractionSlice} might contain a single diffraction pattern, an image of the probe over vacuum, or the average background noise on the detector.
A \texttt{RealSlice} might contain a virtual image, a Boolean mask indicating scan positions to be included or excluded in an analysis routine, or the $x$ and $y$ components of a lattice vector calculated at each scan position.
This last example describes a \texttt{RealSlice} of depth 2, i.e. the data contained in the \texttt{RealSlice} class instance is two distinct 2D arrays ($x$ and $y$ of the lattice vector); in general \texttt{DiffractionSlice} and \texttt{RealSlice} objects can have arbitrary depth.
The \texttt{PointList} class is flexible, containing a set of points of arbitrary length in an arbitrary number of dimensions.
On instantiation of a \texttt{PointList}, a set of coordinates must be specified -- e.g. to specify the positions and intensities of the Bragg disk positions detected in a single diffraction pattern, \texttt{(`qx', `qy', `intensity')} might be used.
Points may then be added or removed from the \texttt{PointList}, e.g. as Bragg disks are detected and then thresholded.
Data in \texttt{PointLists} can be easily extracted or sorted by chosen coordinates.
\texttt{PointListArray} instances are 2D arrays of \texttt{PointLists}, organized in memory to facilitate quick access of the \texttt{PointList} corresponding to a single array element, and are useful when storing a \texttt{PointList} for each scan position.
All these datastructure classes inherit from a parent class called \texttt{DataObject} which facilitates basic searching, storing, and saving functionality for all data generated by py4DSTEM, as well as linking to any relevant metadata.

\subsection{File Structure}
\label{S:filestructure}

py4DSTEM saves data in the HDF5 format, described on the \href{http://www.hdfgroup.org/HDF5/}{HDF5 website} \cite{hdf5}.
A description of the flavor of HDF5 used in py4DSTEM, which we refer to as ``electron microscopy datasets'' or EMD files, is available on the \href{https://emdatasets.com/format/}{EMD website}.
Each HDF5 / EMD file generated by py4DSTEM has a top level group containing all data, allowing for the possibility of nesting many py4DSTEM files in a single, larger file, and version tags to allow for backwards compatibility.
Within the top level group a py4DSTEM file contains 3 high level groups: \texttt{data}, \texttt{metadata}, and \texttt{log}.
The \texttt{data} group typically contains 5 subgroups corresponding to the 5 datastructures discussed in the previous section, and each of these contains any number of subgroups, each storing the contents of a single corresponding dataobject, including its raw data and any relevant metadata (e.g. the length of a \texttt{PointList}, the dimensions of a \texttt{DiffractionSlice}, etc).
This structure makes it possible to bundle all elements of one or more data processing pipelines pertaining to a single raw dataset in a single location, and simplifies reuse between measurements of any shared datastructures.

Loading data necessarily varies based on the input file type.
For its native HDF5 files, py4DSTEM supports scanning the contents of a file before pulling anything into memory, so the entirety of large files need not be loaded if only some subset of smaller dataobjects are required.
For very large datasets, memory mapping of datacubes is supported, whereby the contents of a loaded datacube object are left in non-volatile storage, and individual diffraction pattern are pulled into RAM only as they are accessed, enabling analysis of datasets that are larger than available system RAM. 
Binning during loading is also supported.  
For non-native files, many of the file types used in electron microscopy are proprietary and the contents are not publicly described, which hinders scientific progress within electron microscopy.
py4DSTEM therefore relies on the i/o components of two other open source projects, \href{https://hyperspy.org/}{hyperspy} and \href{https://openncem.readthedocs.io/en/latest/}{openNCEM}.  
Most electron microscopy file formats are currently supported, and to-date the py4DSTEM reader has been tested and works successfully for 4D-STEM data in .dm3, .dm4, etc., formats.


%


The metadata group contains 5 subgroups: \texttt{microscope}, \texttt{sample}, \texttt{user}, \texttt{calibration}, \texttt{comments}, and \texttt{original}.
The \texttt{microscope} group contains information related to the microscope setup and acquisition parameters, such as the accelerating voltage of the beam, the camera length, the convergence angle, and so forth.
The \texttt{sample} group stores information such as the material imaged, synthesis information, and any sample preparation.
The \texttt{user} group is for information related to the scientist or scientists who obtained the data, including names, institutions, and contact information.
The \texttt{calibration} group contains the pixel sizes (in real and diffraction space), as well as any additional calibration information such as rotational offsets, diffraction shifts, and elliptical distortions, which will be discussed in more detail in Sec.~\ref{S:calibration}
The \texttt{comments} group is for any miscellaneous comments.
The \texttt{original} group contains any raw metadata scraped from the original data file.


More details about the program structure, interface, implementation, and usage, including its data handling, modules, the 4D-STEM HDF5 file structure, logging, and metadata handling is available in the py4DSTEM documentation, 
or in the \href{https://github.com/py4dstem/py4DSTEM/}{py4DSTEM repository}.


\section{Basic Processing}
\label{S:basic_processing}

In this section, we discuss the basic processing required for most datasets, namely: preprocessing in Sec.~\ref{S:preprocessing}, Bragg disk detection (for crystalline samples) in Sec.~\ref{S:bragg_disk_detection}, calibration in Sec.~\ref{S:calibration}, polar transformations (for amorphous samples) in Sec.~\ref{S:polar_transformation}, and classification in Sec.~\ref{S:classification}. 
These processing steps are basic in the sense of underpinning all subsequent analyses, rather than in the sense of simplicity;
these methods are not aimed at producing a final measurement or plot, but rather are the necessary preparatory work to ensure such ultimate measurements are possible, and are optimally accurate.  
Measurements and applications are addressed in Sec.~\ref{S:measurements_and_applications}.

\subsection{Preprocessing}
\label{S:preprocessing}

\begin{figure}[htbp]
    \centering
    \includegraphics[width=\columnwidth]{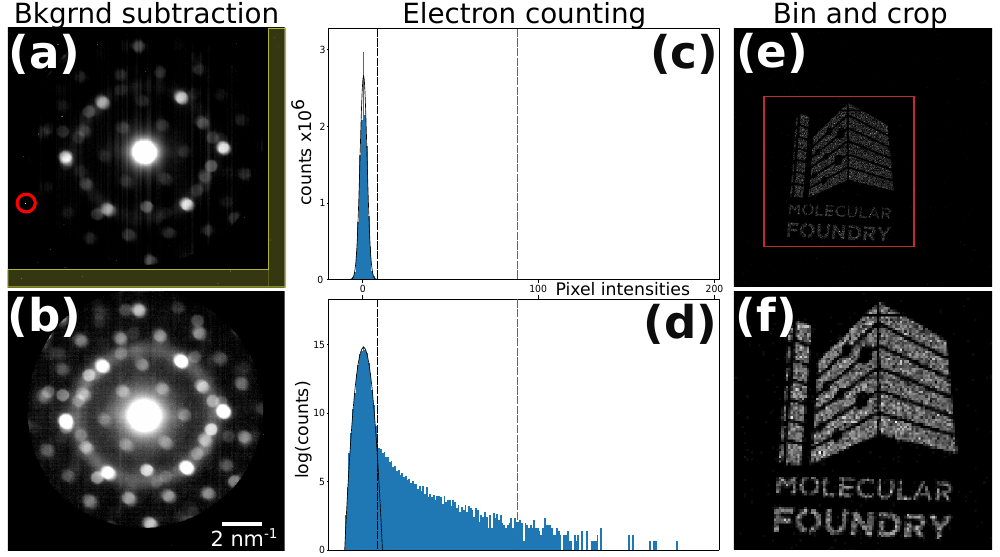}
    \caption{Preprocessing.  (a) A position averaged diffraction pattern of raw 4D-STEM data.  (b) The same position averaged diffraction pattern after subtracting a background determined from the yellow regions in (a).  (c,d)  The initial step of an electron counting procedure, in which minimum and maximum thresholds (black and red dashed lines, respectively) of the pixel intensities are used to rule out background pixels and x-ray strikes.  (e,f) Binning and cropping.}
\label{F:preprocessing}
\end{figure}

This section discusses several preprocessing steps that may be performed on a 4D-STEM dataset.
None of these steps are universally required, however, care in preprocessing can significantly speed up subsequent processing, and lead to higher accuracy and precision in final analyses.

Figure~\ref{F:preprocessing}(a) shows the average of all the diffraction patterns from the GTO dataset.
Vertical streaks are apparent in the image, as well as a handful of individual, erroneously saturated pixels.
Zeroing the hot pixels, calculating a background image and subtracting it from each diffraction pattern, and then calculating a new average diffraction image yields Fig.~\ref{F:preprocessing}(b).
Here, hot pixels were found by detecting outliers in the histogram of pixel intensities across the dataset.
The background was determined by identifying edges of the detector which were beyond the HAADF detector and should ideally have no counts, then using this region (shown in yellow) over many diffraction patterns to calculate the average background streaking.
Alternatively, one or many dark reference images can be recorded directly.
While the nature of noise in a raw 4D-STEM dataset will vary from camera to camera and experiment to experiment, background subtraction is generally recommended.

In 4D-STEM data with a sufficiently low electron dose, it is possible to detect individual electron strike events.
Electron counting, i.e. determining and recording the diffraction space positions of each electron incident on the detector, is beneficial for both noise reduction and data compression.
In py4DSTEM, this is implemented by first calculating a dark reference for the detector.
A histogram of pixel intensity values is then generated from a random sampling of detector frames, and is used to calculate an upper intensity threshold (for excluding x-ray strikes) and a lower intensity threshold (for excluding the background).
In Fig.~\ref{F:preprocessing}, the histograms in panels (c,d) correspond to the low-dose dataset shown in (e,f).
These diffraction patterns were recorded by placing an ``amplitude plate'' aperture in a condenser aperture, as described in \cite{zeltmann2019patterned}.
Looping through each scan position, the dark reference is subtracted and the thresholds are applied to each detector frame, and the local maxima of the resulting image are identified.
These local maxima are considered electron strike events.
Optionally, their positions can be refined to subpixel precision.
The electron counting shown in the figure compresses this data by a factor of $\sim$6000.

The most basic preprocessing functions include reshaping, binning, and cropping data.
Binning and cropping can be performed in either real or diffraction space, and allow large datasets to be reduced to more manageable sizes.
For selected file formats, py4DSTEM also supports data binning on import. 
Figure~\ref{F:preprocessing}e,f shows an electron beam which has been shaped using a structured condenser aperture; from panel (e) to (f) this data has been cropped and binned by a factor of three.
Reshaping the data may be necessary in some cases, for instance, some file formats do not contain complete information about the real space scan shape, and thus can be initially loaded as 3D arrays (with the two real space dimensions collapsed into one) before being correctly reshaped into 4D arrays.

\subsection{Bragg disk detection}
\label{S:bragg_disk_detection}

\begin{figure*}[htbp]
    \centering
    \includegraphics[width=\textwidth]{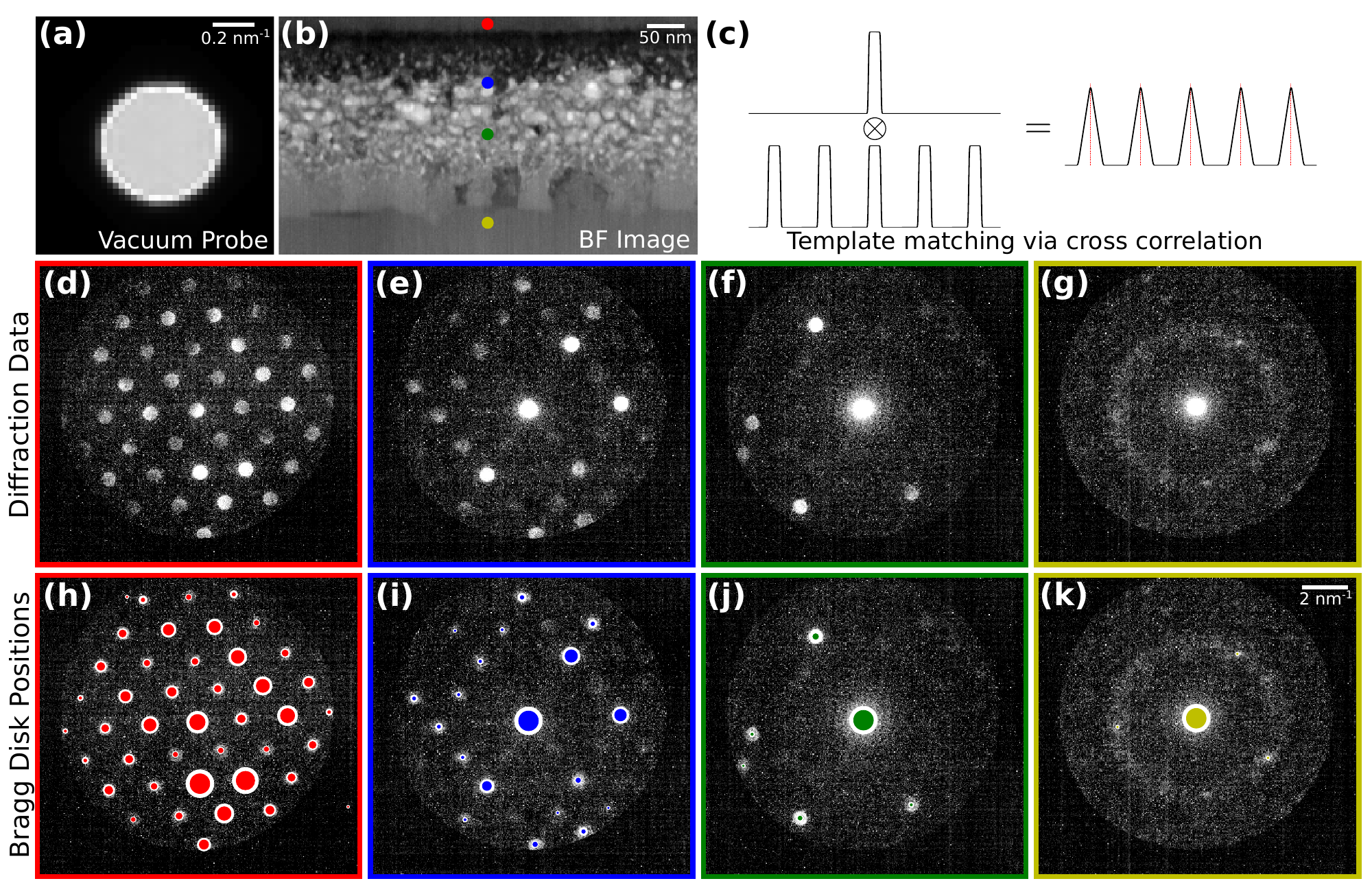}
    \caption{Bragg disk detection in GTO. (a) The vacuum probe.  (b) A virtual bright field image.  (c) Disk detection is accomplished by cross correlation of the probe template with each diffraction pattern.  (d-g) Diffraction patterns corresponding to the four scan positions indicated in (b).  (h-k)  The detected Bragg peaks for these four diffraction patterns.  The size of each circle indicates the cross correlation intensity, a rough approximation for disk intensity.}
    \label{F:diskdetection}
\end{figure*}

For crystalline or semi-crystalline data, analysis generally begins by identifying the locations of all the Bragg disk reflections in each diffraction pattern, which correspond to the reciprocal lattice points of the crystal.
In py4DSTEM, we find the Bragg disk positions in two steps: first, we extract the structure of the probe over vacuum in diffraction space to use as a template.
We then find the Bragg disks by determining all the positions in each diffraction pattern that match the structure of this template \cite{pekin2017strain}.


py4DSTEM includes 3 methods for generating vacuum probes. 
Ideally, we use an image or averaged image stack of the probe over vacuum.
Alternatively, if an experimental 4D-STEM scan contains a vacuum region, or a region with only very thin material (e.g. amorphous carbon support), this can be used to generate a vacuum probe.
In this case, the probes from each vacuum scan position should be aligned, to correct the translation of the diffraction patterns as the beam is scanned, and then averaged.
Finally, if neither of these options is possible, a synthetic probe can be generated.

Once a vacuum probe has been obtained, two additional processing steps are applied, with the purpose of generating a kernel for cross correlative template matching with the individual diffraction patterns.
First, the center of the unscattered electron probe is found and shifted to the origin.
Without this step, all measurements will have an offset, leading to incorrect results.
Second, a Gaussian wider than the probe is subtracted, leading to a region of negative intensity surrounding the probe itself, such that the total integrated intensity of the kernel is zero.
This has two advantages: first, it ensures that the cross correlation of noisy data is, on average, zero where there are no Bragg disks.
Second, the negative kernel intensity penalizes cross correlation values where a Bragg disk and a template are slightly misaligned, enhancing the detectability of correlation maxima where disk/template alignment is perfect.
While subtraction of a Gaussian is a useful heuristic, other approaches are possible, for instance those described in \cite{williamson2015quantitative, pekin2017strain, grieb2017optimization, grieb2018strain, mahr2019influence, padgett2019exit}.
Adding structure to the electron probes using an amplitude mask in the objective aperture has also been shown to significantly enhance the precision of Bragg disk detection \cite{zeltmann2019patterned}.


The Bragg disks are located by calculating the cross correlation of the probe kernel with each diffraction pattern, and then locating the correlation maxima. 
The disk positions can be located with subpixel precision via local Fourier upsampling in the region about each maximum \cite{Guizar-Sicairos:08,Soummer:07}.
py4DSTEM allows for standard cross correlations, as well as phase or hybrid correlations, to be performed at this stage; see Appendix~\ref{A:cc} for detailed discussion.


The detected Bragg disks in each diffraction pattern are a stored in a \texttt{PointList} instance with three coordinates specifying the disk position in the diffraction plane and its cross correlation intensity, $(q_x, q_y, I)$.
The Bragg disks from the complete datacube are stored in a \texttt{PointListArray} instance, with one such \texttt{PointList} for each scan position.
For many analyses, such as strain or orientation mapping, all subsequent computation can be performed on this \texttt{PointListArray} alone, as it contains the most crucial scattering information.
The data compression here is significant, as only 3 numbers are now required to store each Bragg disk.
For a datacube consisting of 512$\times512$ pixel diffraction patterns with a bit depth of 16, 20 detected disks in an average diffraction pattern, and using 64-bit floating point numbers for the disk coordinates, this scheme compresses the data by a factor of approximately 1000.

\begin{figure}[htbp]
    \centering
    \includegraphics[width=\columnwidth]{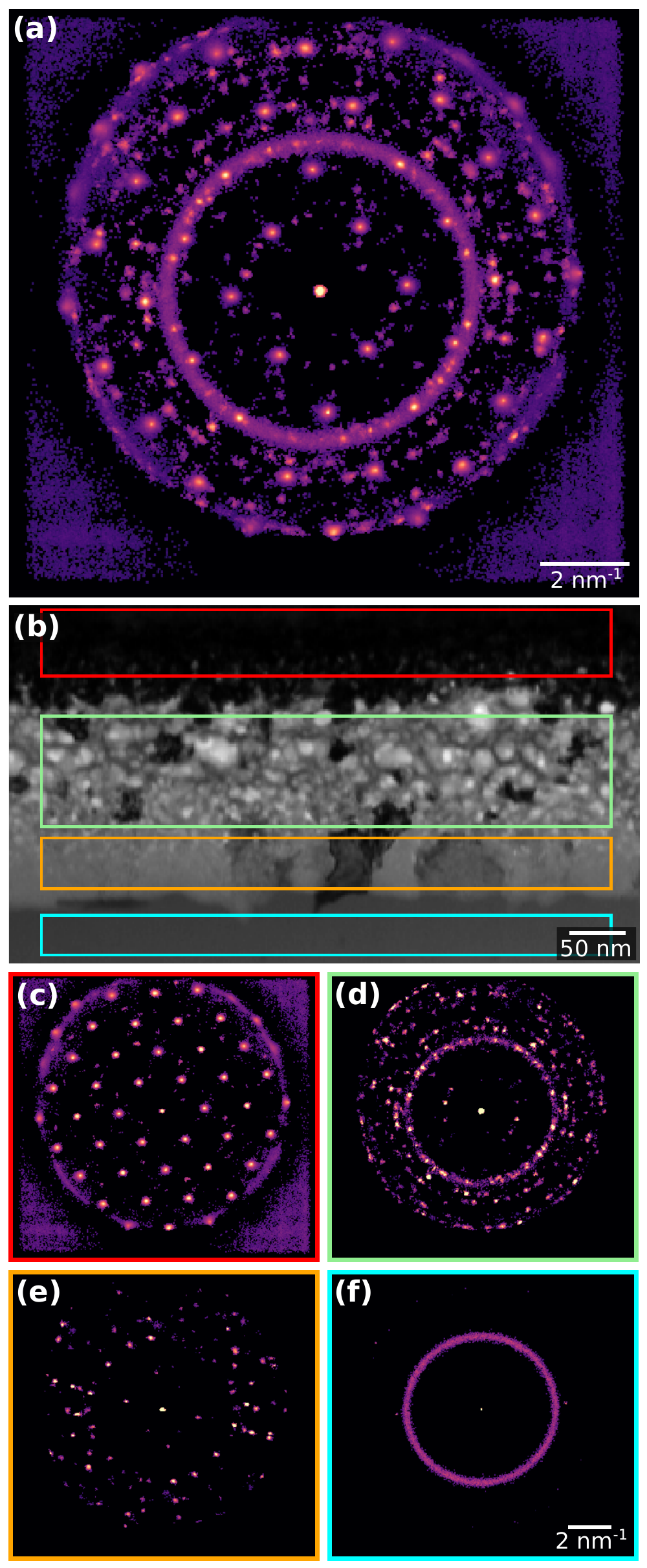}
    \caption{Bragg vector maps of GTO.  (a) The Bragg vector map from the complete dataset.  (b) Virtual bright-field image with boxes indicating four regions of interest.  (d-f) Bragg vector maps generated from the four corresponding regions shown in (b).}
    \label{F:braggvectormap}
\end{figure}

Once the Bragg disks have been detected, all peaks from all scan positions may be collapsed into a single image in the shape of the diffraction plane.
The resulting object is roughly interpretable as a position averaged probability distribution of reciprocal lattice points, and is defined carefully in Appendix~\ref{A:bvm}.
Figure~\ref{F:braggvectormap} shows an example using the GTO dataset.
We refer to this object as a Bragg vector map (BVM).
Figure~\ref{F:braggvectormap}a shows the BVM of the complete GTO 4D-STEM scan, while Figs.~\ref{F:braggvectormap}(c-f) show the BVMs generated from subsets of the scan region indicated in the virtual image shown in Fig.~\ref{F:braggvectormap}b.
The BVM of the single crystal region, Fig.~\ref{F:braggvectormap}c, shows sharp reciprocal lattice peaks of the orthorhombic crystal in the $\langle 01\bar{1}\rangle$ projection.
The BVM of Fig.~\ref{F:braggvectormap}d also contains sharp peaks, now oriented isotropically about the origin, indicating many small, randomly oriented crystal grains.
Figure~\ref{F:braggvectormap}d also shows a faint ring resulting from amorphous scattering in this mixed cystalline/amorphous region.
Note that ideally the BVM would be insensitive to amorphous scattering because it should only contain counts where Bragg scattering occurs, however, false-positive Bragg disk detection can occur in the amorphous halo, resulting the ring here as well as in Fig.~\ref{F:braggvectormap}f.
Figure~\ref{F:braggvectormap}e shows little amorphous signal, sharp peaks indicating crystal scattering, and fewer peaks than in Fig.~\ref{F:braggvectormap}d, suggesting this layer of the sample may contain fewer, larger crystallites.
Figure~\ref{F:braggvectormap}f shows little or no crystalline signal, suggesting a purely amorphous layer.
Phase mapping, found in Sec.~\ref{S:phase_mapping}, confirms these hypotheses about the sample structure.

BVMs are a useful tool in 4D-STEM data processing. 
In py4DSTEM they are used in processing pipelines including calibration (see Fig.~\ref{F:calibration}), classification (see Fig.~\ref{F:classification}), strain mapping (see Fig.~\ref{F:strainmapping}), and others.

\subsection{Calibration}
\label{S:calibration}

Calibration is the single most important step of any quantitatively meaningful 4D-STEM data analysis, as all subsequent measurements hinge on the accuracy of the calibration.
In 4D-STEM a number of calibrations are desirable.
These include correcting shifts of the diffraction pattern from the raster of the beam, correcting elliptical distortions of the diffraction patterns, calibrating the rotational offset between real and diffraction space, and calibrating the pixel sizes.
Which calibrations are required will generally depend on the sample being imaged, the measurements being made, and the required precision.

The data required to perform calibrations is similarly contingent, and depends on the structure of the sample, and which calibrations need to be performed.
An image or a stack of images of the STEM probe over vacuum should always be acquired, and is important for analyses including Bragg disk detection, calibration of the convergence semiangle, and deconvolution of the probe.
Scanning a standard calibration sample of known structure at the beginning or end of a microscope session is highly recommended, and will typically ensure the most accurate calibration of pixel sizes.
Using a standard calibration sample which is polycrystalline is also highly recommended, to facilitate calibration of inevitable elliptical stretching of the diffraction patterns due to imperfect optics and alignments \cite{mahr2019influence}.
Obtaining an image of the probe, positioned over the sample and then highly defocused to create a shadow image in the diffraction plane, is the recommended data for calibrating the real/diffraction space rotational offset.
In some cases it is possible to obtain the necessary calibrations directly from the experimental 4D-STEM scan, however, this is not guaranteed to be possible, and is especially dubious for samples of unknown structure.

\begin{figure*}[htbp]
    \centering
    \includegraphics[width=0.7\textwidth]{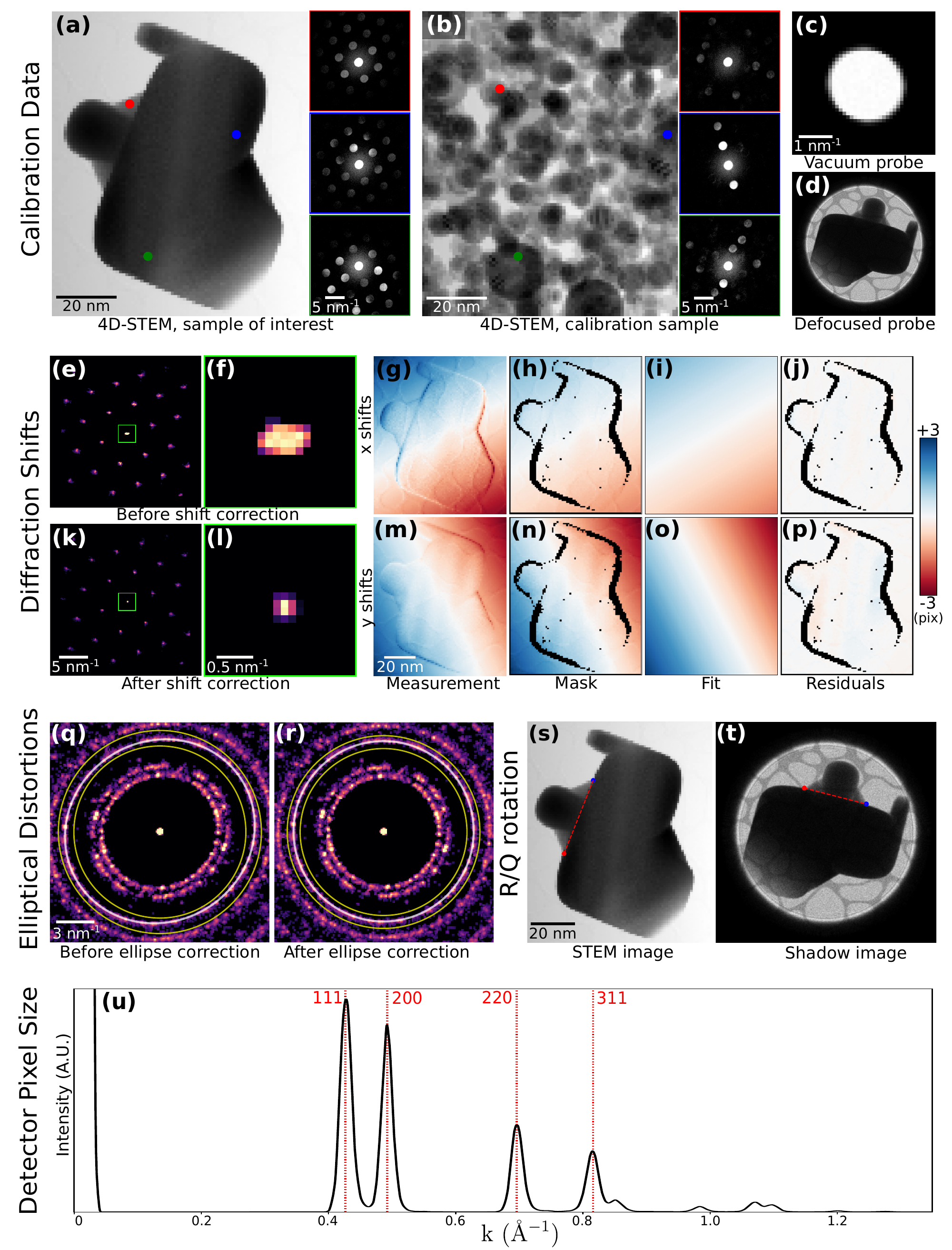}
    \caption{Calibration.  (a-d) The recommended data to collect in order to fully calibrate a 4D-STEM dataset.  Data shown here has been simulated.  (a) A 4D-STEM dataset of a sample of interest, here a strained, single-crystal gold nanoparticle. (b) A 4D-STEM dataset of a standard calibration sample, here a distribution of gold nanoparticles.  In both (a) and (b), a virtual bright-field image and three selected diffraction patterns of the 4D-STEM datasets are shown.  (c) An image or image stack of the probe over vacuum.  (d) An image of the probe over the sample and defocused until a shadow image is visible.  (e-u) A complete set of 4D-STEM calibrations.  (e-p) Measurement and correction of translations of the diffraction patterns with the beam raster.  (q-r) Measurement and correction of elliptical distortions.  (s,t) Measurement of the rotational offset of the electron beam between the real and diffraction planes.  (u) Measurement of the detector pixel size.}
    \label{F:calibration}
\end{figure*}

Figure~\ref{F:calibration} shows the complete calibration of a simulated dataset \cite{ophus2020_4DSTEMcalibrationdata}.
The top row of subpanels shows the data: 
a 4D-STEM scan of the sample of interest, in this case a single-crystalline, strained gold nanoparticle (Fig.~\ref{F:calibration}a);
a 4D-STEM scan of a standard calibration sample, in this case polycrystalline gold (Fig.~\ref{F:calibration}b);
an image of the STEM probe in the diffraction plane (Fig.~\ref{F:calibration}c);
and an image of the STEM probe after defocusing to make a shadow image (Fig.~\ref{F:calibration}d).

Diffraction shifts -- overall translation of the diffraction patterns resulting from the scanning of the electron beam -- yield apparent shifts of the position optic axis from one diffraction pattern to the next \cite{craven1981design}.
The size of the diffraction shifts depend on the real space field-of-view of the scan, on the camera length, and on the particular instrument used; generally speaking, we recommend measuring diffraction shifts in scans larger than a few tens of nanometers, and then applying corrections if deemed necessary.
In py4DSTEM, this calibration is performed by identifying the unscattered beam at each scan position, and measuring the shifts in its position.
These shifts are then fit to a plane or low order polynomial, which can be used to correct the diffraction shifts.
For correcting the shifts, it is possible to shift each diffraction pattern by the measured amount to generate a new, corrected datacube, however, this is slow, resource intensive, and often unnecessary.
Instead, it is often possible to simply use the measured shift values to set the origin of coordinates in any subsequent measurements made on individual diffraction patterns.
Figure~\ref{F:calibration}e-p shows BVMs before (e,f) and after (k,l) diffraction shift corrections have been applied to the measured bragg peak positions.
The zoomed in images centered on the central peak (f,l) illustrate that the blurred peak of (f) collapses to a sharp peak in (l) after shift correction.
In (g-p) we show the initial measurement of shifts of the central disk, a masking step to ignore some subset of data points, a smooth fit to the data, and the residuals, which are all much less than a single pixel.

Elliptical distortions, in which circular features about the optic axis are stretched into ellipses, are generally experimentally unavoidable \cite{mahr2019influence}.
These result from imperfect alignments, including off-axis illumination on the probe-forming condenser aperture, stigmation in the post-specimen optics, and finite tilt of the detector plane relative to the plane normal to the optic axis.
Even in a well aligned system these distortions may be significant, and are therefore important to correct in many quantitatively sensitive experiments.
In py4DSTEM, elliptical distortions can be measured by fitting an elliptical function to data within some specified annular region, as shown in Fig.~\ref{F:calibration}(q,r).
The functional forms of the fits are discussed in more detail in Appendix~\ref{A:ellipses}.
With elliptical fits in hand, the elliptical distortions can be corrected.
For crystalline data in which the Bragg peaks have been measured and subsequent analysis will be performed on the measured peak positions only, correction may be accomplished by shifting the peak positions while leaving the raw data untouched.
Figure~\ref{F:calibration}r shows a BVM after such correction has been performed.
An alternate approach to elliptical correction is to take a polar-elliptical transform, effectively re-sampling the data into a coordinate system which shares the data's ellipticity.
This latter approach is frequently useful in analysis of amorphous datasets, and is discussed further in Sec.~\ref{S:polar_transformation}.

There is in general some angle of rotation between the electron beam in the sample plane and in the detector plane.
Thus in order to correctly map orientations measured in the diffraction plane into real space, it is necessary to measure and account for this rotational offset.
The simplest and most robust way to measure the offset is to compare a STEM image to a defocused probe shadow image.
Any STEM image will suffice, provided that the same features are visible in the STEM and shadow images, and in Fig.~\ref{F:calibration} the bright field virtual image is used.
Two identical points in each of the two images are identified in Fig.~\ref{F:calibration}s,t and are then used to calculate the rotational offset.
If a shadow image has not been obtained, other methods to determine the rotational offset are possible, however, will necessarily be less robust.
Two additional techniques for rotational calibration are provided in py4DSTEM, both based on the principles of differential phase contrast imaging.
As a result, these methods tend to work well when the assumptions of differential phase contrast hold.
They are discussed further, along with the relevant caveats, in Sec.~\ref{S:DPC}.

Calibration of the diffraction space pixel size minimally requires measuring a single diffraction vector with a known spacing.
More accurate measurement is possible by fitting to several known spacings.
Figure~\ref{F:calibration}u shows a radial integral (see Sec.~\ref{S:polar_transformation}) of the elliptically corrected Bragg vector map shown in Fig.~\ref{F:calibration}r.
By indexing the peaks observed and using the known lattice spacing of gold, we use the measured peak positions to calculate the detector pixels size.
The horizontal axis of these plots can then be written in physical units of \AA$^{-1}$.

Ideally, the real space pixel size is determined by the distance the electron probe is rastered by the scan coils between detector frames, and is therefore equivalent to the size calibration of the instrument's STEM scan.
For this reason, processing tools for re-calibration of the real space pixel size are not provided.
However, should such calibration be desired, it is straightforward to edit the py4DSTEM metadata based on independent measurement of the real space pixel sizes.
When specimen drift leads to large deviations of the pixel size and scan direction angles, further pixel size measurements and drift correction may be required \cite{sang2014revolving, ophus2016correcting, savitzky2018image, wang2018correcting}.

\subsection{Polar Transformation}
\label{S:polar_transformation}


\begin{figure}[htbp]
    \centering
    \includegraphics[width=0.95\columnwidth]{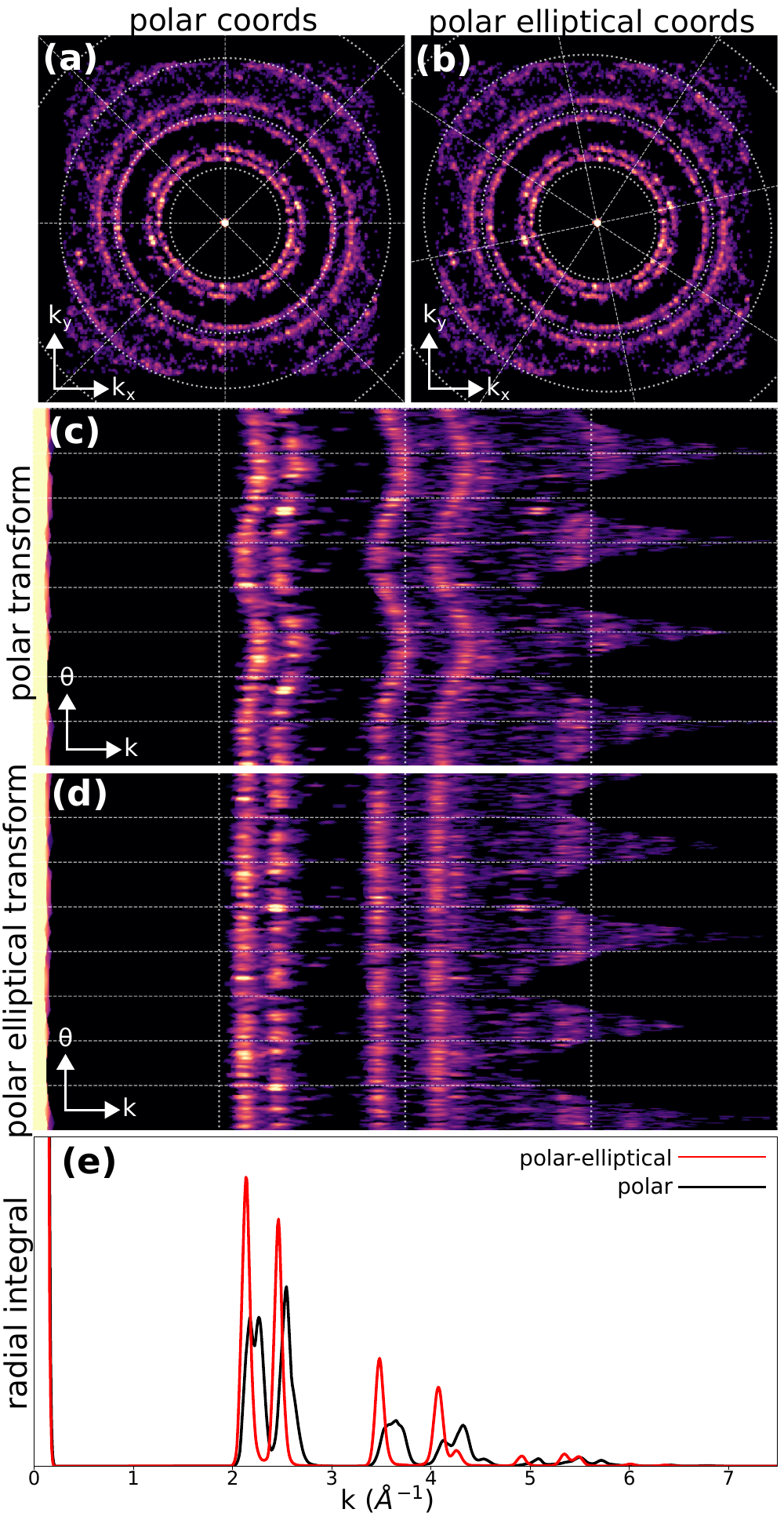}
    \caption{Polar and polar-elliptical transforms.  (a) Bragg vector map of the simulated gold nanoparticle calibration dataset, overlaid with a polar coordinate grid.  (b) Identical data to (a), but overlaid with a polar-elliptical grid calibrated to this data.  (c) The polar transform corresponding to (a).  Note that the rings have been mapped to sinusoids, due to elliptical distortions.  (d) The polar-elliptical transform corresponding to (b).  The rings now map to lines, indicating that the elliptical calibration is correct.  (e) Radial integrals calculated from the polar (\textit{red}) and polar-elliptical (\textit{black}) transforms.}
    \label{F:polar_transform}
\end{figure}

Transformation from Cartesian to polar coordinates is an important operation in many 4D-STEM analyses, especially of amorphous data.
Sections~\ref{S:RDF} and \ref{S:FEM} discuss two examples, fluctuation electron microscopy and radial distribution function analysis.
Polar-elliptical transformations are useful for correcting elliptical distortions, as discussed in Sec.~\ref{S:calibration}.
This also enables calculation of elliptically corrected radial integrals.

Figure~\ref{F:polar_transform} shows the transformation of the BVM of the simulated calibration sample of gold nanoparticles described in Sec.~\ref{S:calibration}.
Both a polar (a,c) and polar-elliptical (b,d) transformation have been performed, in the latter case using elliptical parameters fit from the image.
In the polar case we see that, just as the circular coordinate axes poorly align with the data in (a), so too do the rings turn into vertical sinusoids in (c).
In contrast, in (b) the axes and data are well aligned, and in (d) the rings turn into vertical lines rather than sine curves.

Radial integration of a single or averaged diffraction pattern is an important operation, providing higher SNR information about electron scattering at each spatial frequency, at the expense of losing any orientation information.
The polar-elliptical transform makes elliptically corrected radial integration easy -- just sum along the theta axis of the transformed data.
Figure~\ref{F:polar_transform}e shows an example, with the radial integral calculated from the calibrated polar-elliptical transform in black and the radial integral from the simple polar transform in red.
Note that the simple radial integral broadens peaks and, in the case of the first peak, splits a single peak into two apparent, but spurious, peaks.

\subsection{Classification}
\label{S:classification}

\begin{figure*}[htbp]
    \centering
    \includegraphics[width=\textwidth]{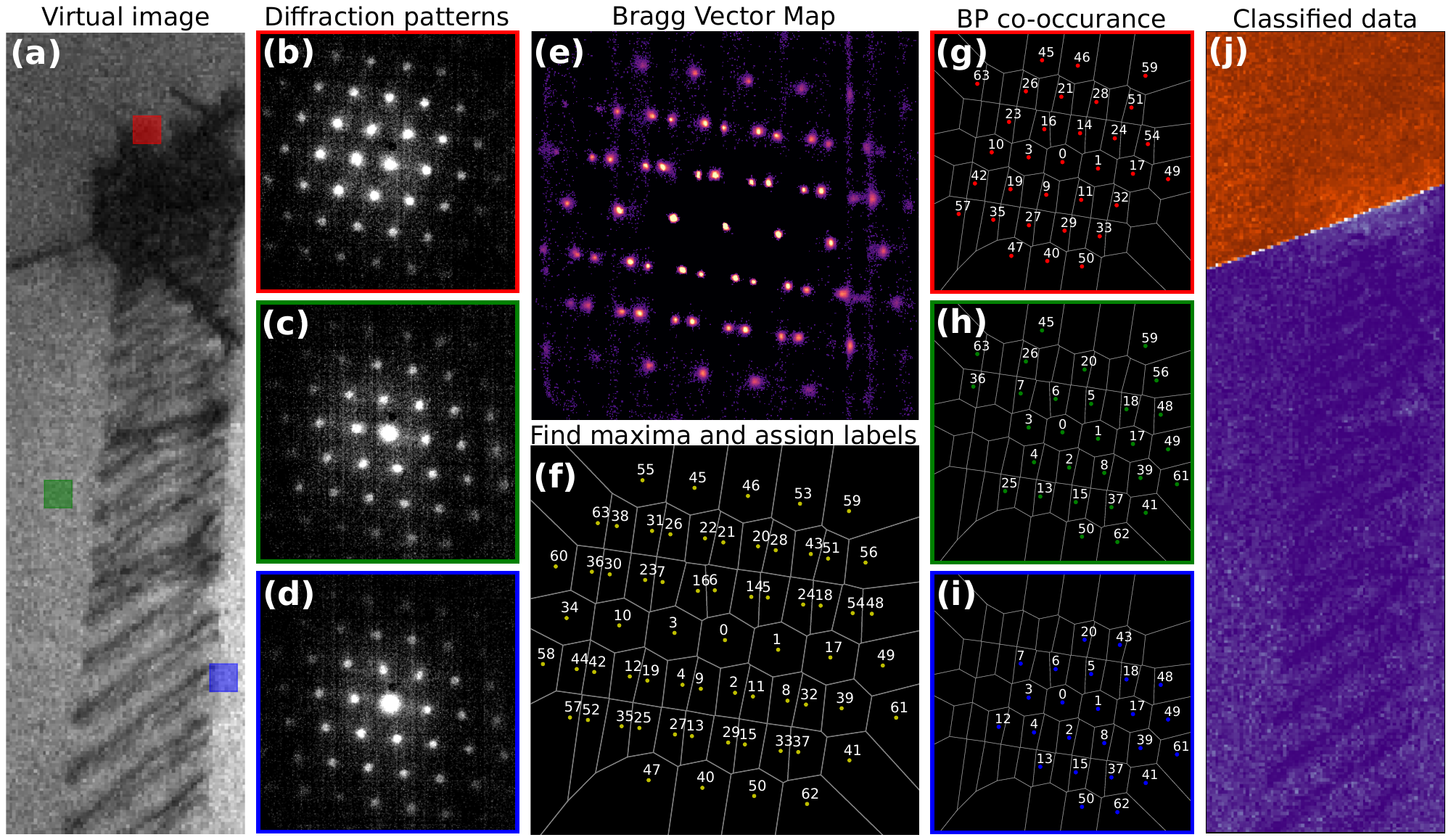}
    \caption{A 4D-STEM classification algorithm. (a) Virtual image of a 4D-STEM dataset of a twin boundary. (b-d) Averages of 100 diffraction patterns each from the regions shown in (a).  (e) The Bragg vector map.  (f) BVM maxima have been located, labelled, and used to segment the diffraction plane.  (g-h) The segmentation in (f) is used to label the Bragg peaks in each diffraction pattern.  (j) Co-occurance of Bragg peaks is used as a criterion to assign scan positions to classes, resulting in a classification which clearly identifies the twin boundary.}
    \label{F:classification}
\end{figure*}

In the context of 4D-STEM, classification refers to assigning one or more integer values to each scan position, which identify this position with associated classes.
Ideally, each class corresponds to a type of diffraction pattern, or to structurally meaningful features or motifs, such that a scan position will be included in a given class if and only if its diffraction pattern contains these features.
Virtual imaging, and thoughtful combination of virtual images and colormaps, is often the easiest way to visually differentiate distinct structural regions, and can be a powerful tool for microanalysis \cite{tao2009_SEND, gammer2015diffraction, zhang2017electroplating, shukla2018effect}.
By identifying each pixel with discrete class types classification goes a step further, facilitating subsequent analyses as well as enabling generation and identification of class diffraction patterns \cite{brunetti2011confirmation, gallagher2019nanoscale}.

Figure~\ref{F:classification} shows a simple classification example.
A 4D-STEM scan was taken of a medium entropy alloy 
containing a twin boundary, which is about three quarters of the way up the virtual image in Fig.~\ref{F:classification}a.
Diffraction patterns averaged from 100 scan positions each about the positions shown with red, green, and blue sqaures are shown in Fig.~\ref{F:classification}b-d.
Inspection reveals that the reciprocal lattice in Fig.~\ref{F:classification}b is twinned with respect to that of Figs~\ref{F:classification}c,d.
This dataset is therefore an excellent testbed for a classification algorithm because the correct answer is immediately apparent: each diffraction pattern in this dataset should be assigned to one of two classes, according to the side of the twin boundary where it falls.

The algorithm proceeds as follows.
First, all Bragg disks are located, as described in Sec.~\ref{S:bragg_disk_detection}.
Next, the BVM is calculated, after any relevant calibrations such as diffraction shift correction have been performed -- see  Fig.~\ref{F:classification}e.
The $N$ maxima of the BVM are then located.
A Voronoi tesselation of the diffraction plane is constructed using these maxima as the initial points, which carves the diffraction plane into a set of $N$ regions, each of which is defined as the set of all points closest to one BVM maximum \cite{barber1996quickhull} - see Fig.~\ref{F:classification}f.
Each of these $N$ regions is assigned an integer value.
Next, the set of Bragg peaks which has been detected at each scan position is retrieved, and each peak is assigned a label according to which Voronoi region it falls in - see Fig.~\ref{F:classification}g-i.
At this stage the complexity of the data has been reduced significantly - for each scan position, we have a small set of integers encoding where Bragg scattering occurred, rather than an entire 2D diffraction pattern.
Initial classes are identified by determining which Bragg peaks co-occur with the highest frequency, and these classes may then be refined, for instance via non-negative matrix factorization.
Here, the final result is shown in Fig.~\ref{F:classification}j, with the data cleanly separated along the twin boundary.
More detailed discussion of the algorithm can be found in Appendix~\ref{A:classification}, and more complex classification example can be found in Sec.~\ref{S:phase_mapping}.


\section{Measurements and applications}
\label{S:measurements_and_applications}

In this section, we build on the techniques described in Sec.~\ref{S:basic_processing} to make various measurements of physical interest from 4D-STEM datasets.
In Sec.~\ref{S:virtual_imaging} we generate virtual images.
In Sec.~\ref{S:phase_mapping} we apply the classification algorithm discussed in Sec.~\ref{S:classification} to the GTO dataset to retrieve maps of various crystalline and amorphous phases present in the complex, nanostructured sample. 
In Secs.~\ref{S:crystalline_strain_mapping} and~\ref{S:amorphous_strain_mapping} we calculate strain maps from crystalline data, and from amorphous data respectively.
In Secs.~\ref{S:RDF} and~\ref{S:FEM} we further analyze amorphous samples, calculating radial distribution functions in the former section and performing fluctuation electron microscopy analysis in the latter section.
We conclude with two phase retrieval methods for reconstructing the sample potential, demonstrating differential phase contrast imaging in Sec.~\ref{S:DPC} and ptychography in Sec.~\ref{S:ptychography}.

\subsection{Virtual Imaging}
\label{S:virtual_imaging}

\begin{figure*}[htbp]
    \centering
    \includegraphics[width=\textwidth]{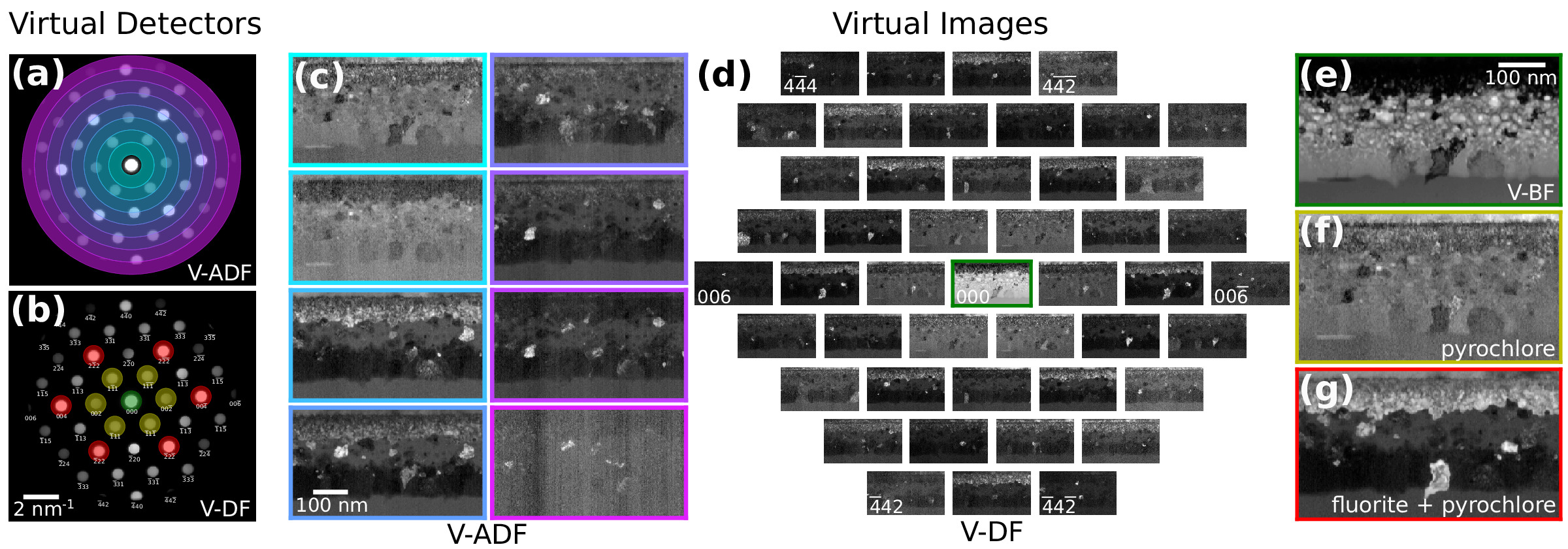}
    \caption{Virtual imaging.  (a) Virtual annular dark-field detectors.  (b) Virtual bright field (\textit{green}) and dark-field (\textit{yellow, red}) detectors.  (c) Virtual annular dark-field images.  (d) Virtual dark-field images corresponding to circular detectors about each of the indexed Bragg peaks.  (e) Virtual bright-field image.  (f,g) Virtual images corresponding to the yellow and red detectors shown in (b), respectively.  The inner, yellow peaks are only present in one of the two expected crystal structures in this system.}
    \label{F:virtualimaging}
\end{figure*}

In a traditional STEM experiment, many imaging modalities are possible, by placing detectors of different geometries in different positions in the diffraction plane \cite{pennycook2011_STEM_textbook}.
4D-STEM enables virtual recreation of a wide swath of such imaging modalities in post processing \cite{zaluzec2003computationally, fundenberger2003polycrystal, lupini2015ptychographic, fatermans2018single}.
See Appendix~\ref{A:STEM}.

Figure~\ref{F:virtualimaging}a,b shows an averaged diffraction pattern from the single crystalline region of the GTO sample, overlaid with various virtual detectors which were used to generate the images in Fig.~\ref{F:virtualimaging}c-g.
Figure~\ref{F:virtualimaging}a shows annular dark field detectors of various inner and outer collection angles, and their corresponding virtual images are shown in Fig.~\ref{F:virtualimaging}c.
The Miller indices of each Bragg reflection in the $\langle 110 \rangle$ projection are shown in Figure~\ref{F:virtualimaging}b, and virtual images corresponding to a detector placed about each of these peaks are shown in Fig.~\ref{F:virtualimaging}d.
Here, a single 4D-STEM scan is used to virtually recreate images analogous to 45 distinct traditional dark-field TEM images, similar to \cite{gammer2015diffraction}.

Figure~\ref{F:virtualimaging}b shows three detectors colored green, yellow, and red, corresponding to the three virtual images shown in Fig.~\ref{F:virtualimaging}e-g.
The first is a virtual bright field image, while the latter two use virtual detectors which would be challenging to realize physically, but which are of particular interest because of the structural significance of the yellow and red peaks to two crystalline phases in this system:
the red peaks are present in both of the two expected single crystal phases (pyrochlore and fluorite), while the yellow peaks vanish in the higher symmetry fluorite phase.
Thus with 4D-STEM, it is possible to virtually recreate images corresponding to every possible integrating STEM detector geometry, and also to generate complex, bespoke detectors matched to the sample structure and properties of interest.

\subsection{Phase Mapping}
\label{S:phase_mapping}

\begin{figure*}
    \centering
    \includegraphics[width=\textwidth]{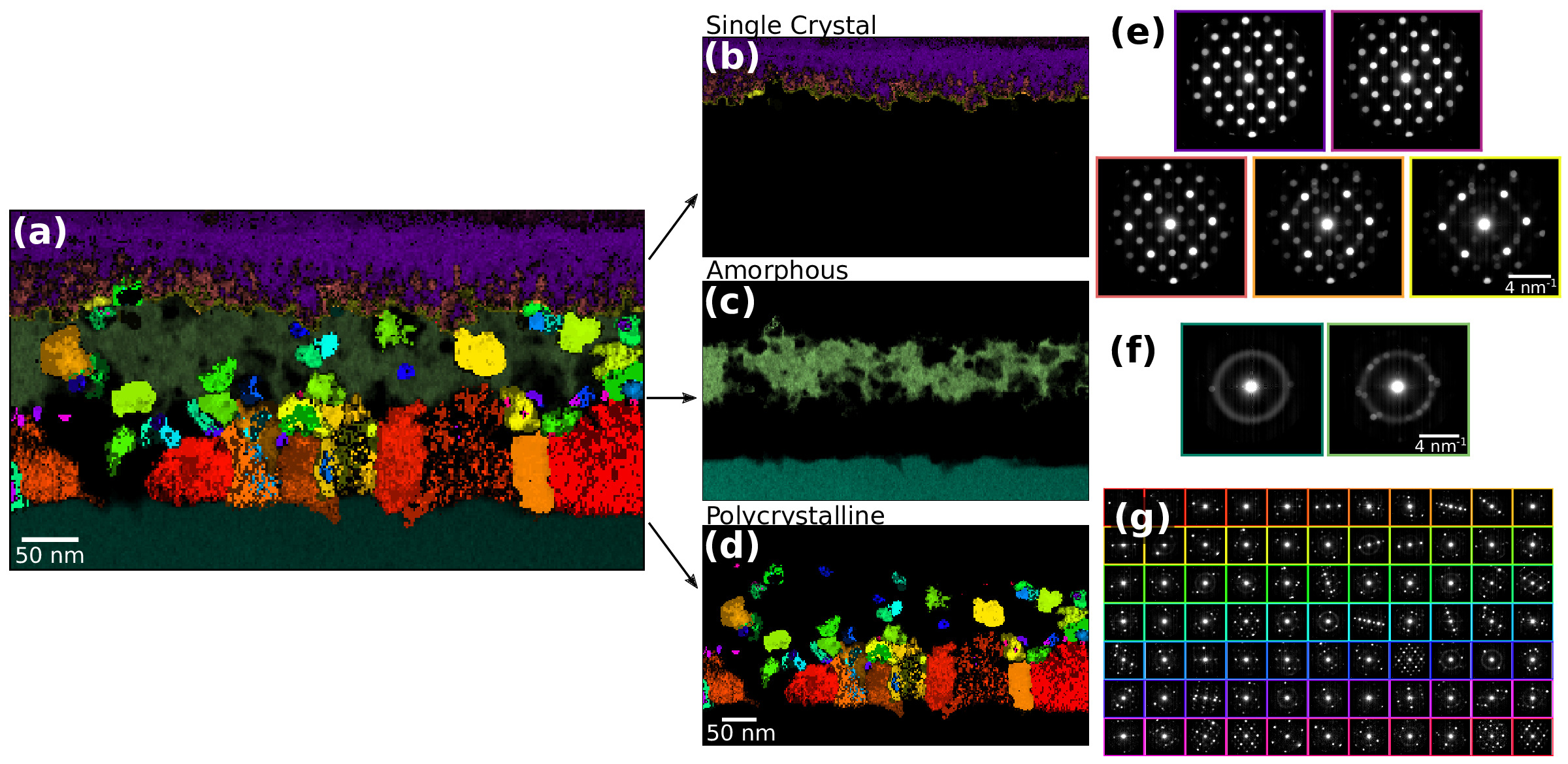}
    \caption{Phase mapping.  (a) All of the structurally distinct phases identified in this system, using the classification algorithm described in \ref{S:classification}.  (b,e) The single crystal phases and their class diffraction patterns.  (c,f) The amorphous phases and their class diffraction patterns.  (d,g) The polycrystalline phases and their class diffraction patterns.}
    \label{F:phasemapping}
\end{figure*}


An important problem in many applications is mapping distinct structural phases, and potentially many phases, present within a single sample \cite{rauch2010automated, brunetti2011confirmation, kobler2013combination, gallagher2019nanoscale}.
In this section we demonstrate mapping regions of a 4D-STEM scan in which the diffraction patterns are sufficiently similar to be considered a single type, using the classification algorithm discussed in Sec.~\ref{S:classification}.
This therefore constitutes `phase' mapping in the sense of distinguishing regions of structural similarity, defined in terms of differences in the measured diffraction patterns.
These differences may result from the presence of distinct crystal structures, crystal grains of various orientations, amorphous regions, and so on.
The meaning of any one of these phases must be interpreted in the context of the particular sample, and the details of each phases' average diffraction pattern  \cite{schwarzer1998automated}.

We return to the GTO dataset as an example.
The results are shown in Fig.~\ref{F:phasemapping}.
The classification algorithm identifies 82 distinct crystalline phases, including 5 single crystal phases (Fig.~\ref{F:phasemapping}b,e) and 77 smaller crystallites (Fig.~\ref{F:phasemapping}d,g).
We then additionally identified two amorphous phases (Fig.~\ref{F:phasemapping}c,f).
This was accomplished by masking away all detected Bragg peaks, calculating radial integrals of the masked diffraction patterns, then using these curves as inputs to a non-negative matrix factorization algorithm. 
Masking Bragg peaks is not required for purely amorphous data, but is essential for mixed amorphous/crystalline specimens, as Bragg scattering even from small crystallites in a primarily amorphous matrix would otherwise dominate the radially integrated signal.

In this dataset, we find a single crystal region which appears to transition smoothly from a pyrochlore structure (Fig.~\ref{S:phase_mapping}b, \textit{dark purple}, and Fig.~\ref{F:phasemapping}e, upper left) to a flourite structure in which the superlattice reflections vanish (Fig.~\ref{S:phase_mapping}b, \textit{yellow}, and Fig.~\ref{F:phasemapping}e, lower right).
Below the single crystal region is a mixed crystalline/amorphous region (Fig.~\ref{S:phase_mapping}c, \textit{lighter green}, and Fig.~\ref{F:phasemapping}f, right).
Below this is a layer of larger crystallites (Fig.~\ref{S:phase_mapping}d,g), followed by a pure amorphous region (Fig.~\ref{S:phase_mapping}c, \textit{darker green}, and Fig.~\ref{F:phasemapping}f, left).
With a phase map in hand, any number of additional analyses, such as the orientation or size distribution of the crystallites, or the strain in the single crystal (see Fig.~\ref{F:strainmapping}), become readily calculable.

\subsection{Crystalline Strain Mapping}
\label{S:crystalline_strain_mapping}

\begin{figure}
    \centering
    \includegraphics[width=\columnwidth]{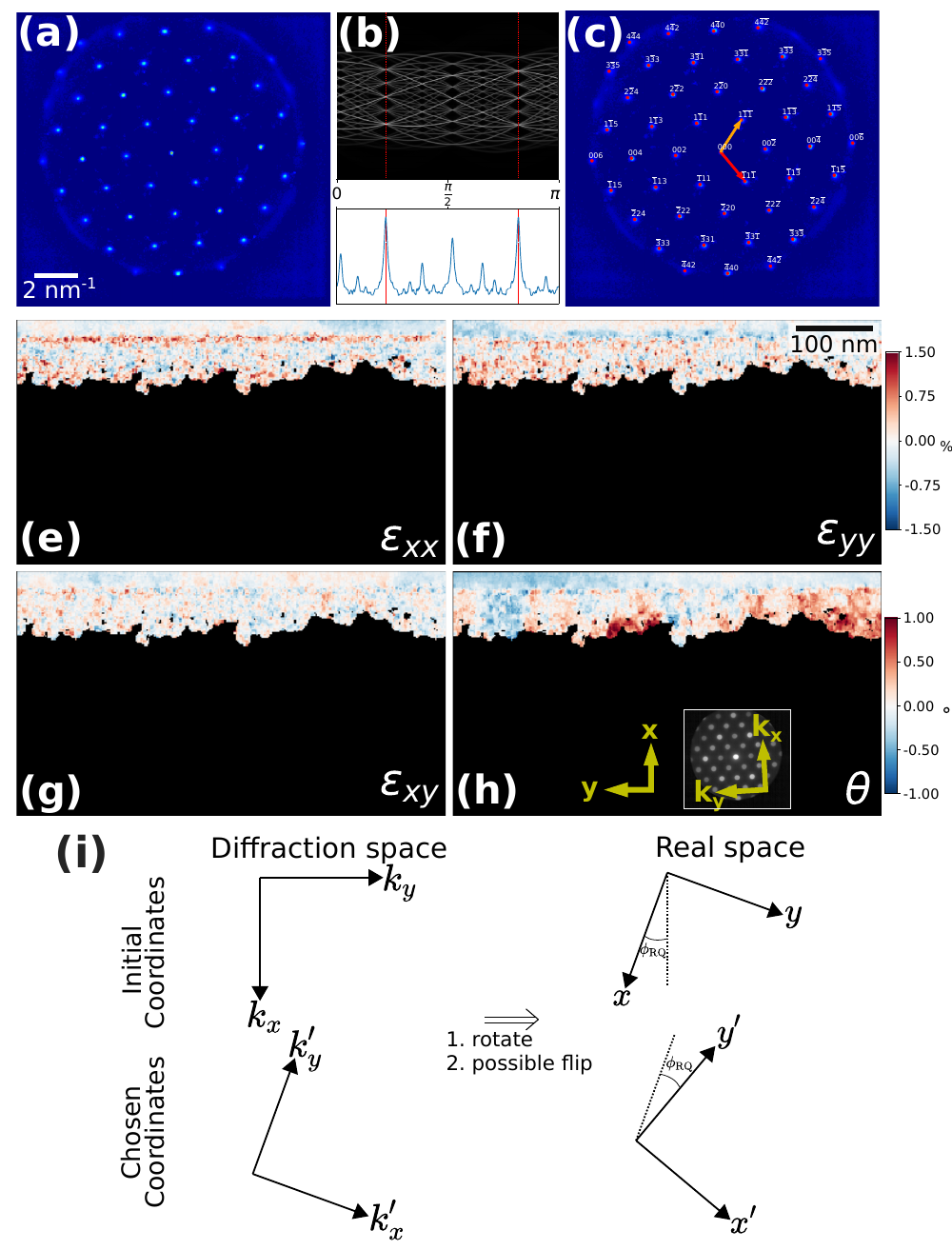}
    \caption{Crystalline strain mapping.  (a) Bragg vector map of the crystalline region of the GTO sample.  (b) Automated detection of the lattice vectors, using the Radon transform.  (c) The indexed Bragg vector map.  (e-h) Strain maps of the single crystal region.  The upper color bar applies to (e-g), and the lower colorbar to (h).  (i) The relevant coordinate systems in real and diffraction space.}
    \label{F:strainmapping}
\end{figure}

The diffraction pattern of a crystalline sample from a low index zone axis contains a grid of Bragg disks given by the reciprocal lattice of the sample. Therefore the spacing of the Bragg disks is inversely proportional to the real space atomic spacing. Precise measurements of the reciprocal lattice vectors can therefore be used to map the local strain present in a crystalline sample, given by the deviations of the lattice from the ideal spacing and angles \cite{usuda2004strain, liu2008strain, beche2009improved, sourty2009using, favia2010nano, uesugi2011evaluation}.

In Fig.~\ref{F:strainmapping} we map the strain of the single crystal regions of the GTO data.
Obtaining a strain map begins with Bragg peak detection as discussed in Sec.~\ref{S:bragg_disk_detection} and data calibration as discussed in Sec.~\ref{S:calibration}.
Beginning from the calibrated BVM of the region of interest (Fig.~\ref{F:strainmapping}a), the average reciprocal lattice vectors are extracted by taking its Radon transform, and then finding the projection angles at which the peaks of the BVM align (Fig.~\ref{F:strainmapping}b).
With the lattice vectors in hand, the BVM peaks are indexed (Fig.~\ref{F:strainmapping}c).
We then refine the reciprocal lattice vectors for each diffraction pattern by performing a fit to its set of detected Bragg peaks, using the average lattice vectors as an initial guess and weighting the fit according to the cross correlation intensities of the detected peaks.
A reference lattice is chosen, and the infinitesimal strain tensor is computed at each beam position by examining the deviation of its local lattice vectors from the reference lattice.
For further discussion see Appendix~\ref{A:strain}.

The results of this analysis are shown in Fig.~\ref{F:strainmapping}e-h.
Here, the $x$- and $y$-directions are shown in both real and diffraction space with red and orange arrows, respectively. 
$\epsilon_{xx}$ and $\epsilon_{yy}$ refer to the compressive/tensile (negative/positive) strain of the lattice along the $x$ and $y$ directions shown, while $\epsilon_{xy}$ and $\theta$ are the shear strain and the rotation of the lattice, respectively.
Among other revealing features, the $\epsilon_{xx}$ map in this data shows significant tensile strain along the interface of the parent and recrystallized single crystal, indicating stretching of the crystal perpendicular to the interface.

The choice of reference lattice is crucial to obtaining meaningful strain maps.
In the simplest case, the experimental 4D-STEM scan contains a region of known undeformed lattice, which can be used directly to define the reference lattice.
Alternatively, it is possible to obtain a separate scan of unstrained material to use as reference, however in this case, good calibrations are essential -- see Sec.~\ref{S:calibration}.
With good calibrations and a known crystal structure, it is also possible to define a reference lattice by hand.
In the case of the GTO dataset, in which there is a parent crystal at the top of the image and a region of recrystallization below, the parent crystal can be used as reference.

Strain tensor values depend, in general, on the choice of coordinate system.
It is therefore necessary to specify coordinates; without this specification, e.g. by including the coordinate axes on the plots, strain maps are not physically interpretable.
Because there is some arbitrary rotation between real and diffraction space in 4D-STEM data, it is also important to show the orientation of the axes in both real and diffraction space.
In Fig.~\ref{F:strainmapping}h, two sets of yellow axes show the chosen coordinate with respect to which the strain maps are are measured, in real space and diffraction space respectively.
In this data, the rotation between the two was small ($\sim2^\circ$), however note that in general it need not be, and will vary between microscopes.
The best coordinate system to use for a give strain map depends on the sample and the relevant material questions.
Typically, orienting one of the principle axes along some important crystallographic direction is best, and in Fig.~\ref{F:strainmapping} the strain $x$-direction has been oriented along the $\langle 1\bar{1}0 \rangle$ direction, which is also direction of ion bombardment and of recrystallization.
In a strain mapping workflow in py4DSTEM, calculating the strain from the reference lattice produces a strain map with respect to a coordinate system oriented along the detector frame (Fig.~\ref{F:strainmapping}i, top row);
typically, some coordinate orientation which is sensible for the system and questions under study should then be chosen, and the strain map rotated into this coordinate system (Fig.~\ref{F:strainmapping}i, bottom row).

\subsection{Amorphous Strain Mapping}
\label{S:amorphous_strain_mapping}

Electron diffraction experiments of amorphous materials, or materials containing a substantial fraction of an amorphous phase, will typically include ring-like features with a radius given by a characteristic scattering length.
Similarly to crystalline materials, a local increase or decrease in the average atomic spacing (i.e. strain) in amorphous materials will cause a decrease or increase respectively in the amorphous ring radius.
By fitting an elliptical function to each diffraction image, we can directly measure these deviations due to local strain.
This has been demonstrated both in individual TEM diffraction images \cite{ebner2016local} and in \emph{in situ} 4D-STEM experiments \cite{gammer2018local}.

\begin{figure}
    \centering
    \includegraphics[width=\columnwidth]{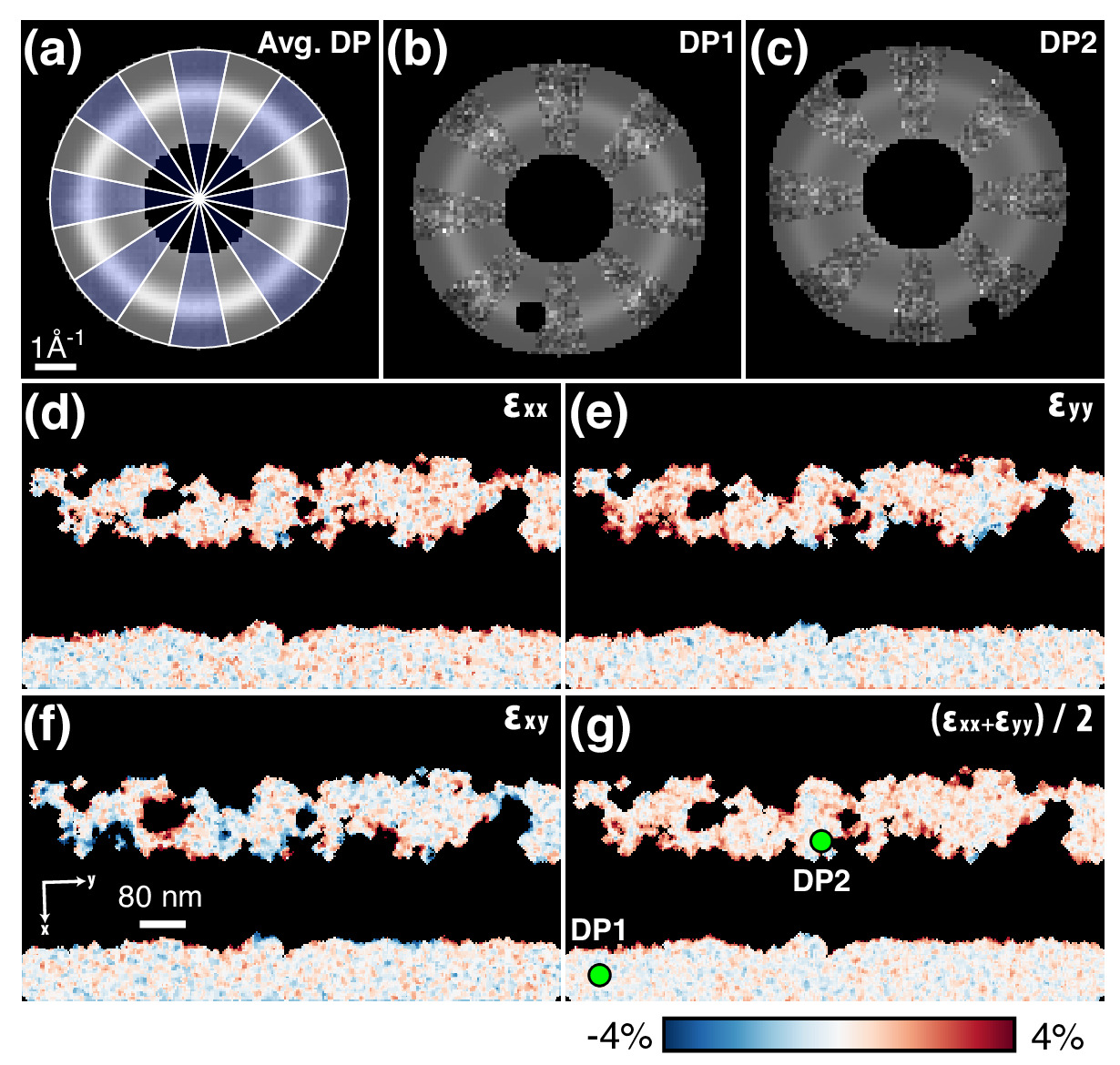}
    \caption{Amorphous strain. (a-c) Elliptical fits to the average amorphous diffraction pattern, and two two selected diffraction patterns.  In (a), blue wedges show the data, while clear wedges show the fit function.  In (b) and (c), the data and the fit are similarly interleaved, and Bragg scattering has been masked away to ensure good fitting.  (d-e) The compressive/tensile strains $\epsilon_{xx}$ and $\epsilon_{yy}$, the shear strain $\epsilon_{xy}$, and the dilation $\frac{1}{2}(\epsilon_{xx}+\epsilon_{yy})$.}
    \label{F:amorphousstrain}
\end{figure}

In py4DSTEM, we have implemented strain measurements of amorphous materials using the same elliptic fitting routines described in Sec.~\ref{S:calibration} and Appendix~\ref{A:ellipses}.
Figure~\ref{F:amorphousstrain}a-c show the elliptical fits.
In each of the three plots shown, the data being displayed alternates in a pinwheel pattern between the data and the fit, for easy visual assessment of the fit quality.
In the average diffraction pattern of the pure amorphous region, Fig.~\ref{F:amorphousstrain}a, the data (shaded blue) is in excellent agreement with the fit.
Using this fit as an initial guess, noisier individual diffraction patterns like Fig.~\ref{F:amorphousstrain}b,c can then be fit as well.
In data containing mixed amorphous and crystalline material, to obtain good elliptical fits to the amorphous signal it is important to mask off any Bragg scattering.
In Fig.~\ref{F:amorphousstrain}b,c the smaller black circles represent such masked regions.

Figure~\ref{F:amorphousstrain}d-g shows the strains computed beginning from these fits, then finding the deviation of the elliptical distortions from a reference.
Here the median of the fully amorphous region is used.
As with crystalline strain mapping, the choice of reference is important, and should be selected carefully based on the individual experiment.
Figure~\ref{F:amorphousstrain}d-f, showing the compressive/tensile strains along the shown $x$ and $y$ directions as well as the shear strain, are comparable to the crystalline $\epsilon_{xx}$, $\epsilon_{yy}$, and $\epsilon_{xy}$ plots from Fig.~\ref{F:strainmapping}.
Figure~\ref{F:amorphousstrain}g additionally shows $\frac{1}{2}(\epsilon_{xx}+\epsilon_{yy})$, representing the local dilation of the structure.
Across the four shown amorphous strain plots we observe local structural changes, especially at the crystalline-amorphous interfaces. 


\subsection{Radial Distribution Functions}
\label{S:RDF}

The radial distribution function (RDF), or $g(r)$, describes the relative density of atoms some distance $r$ from a given atomic position.
Thus the RDF characterizes the distribution of distances between atoms in a given material.
It can serve as an important fingerprint for amorphous materials, as it gives information about the distance and density of neighboring shells of atoms, which depend on the material's structure, chemistry and defect density \cite{srolovitz1981radial}.
In this section, we qualitatively discuss the calculation of the RDF, and the structure of the resulting plot.
Formal discussion of our methods, which follow \cite{mu2016radial, mitchell2012rdftools}, are found in Appendix~\ref{A:RDF}.

\begin{figure}
    \centering
    \includegraphics[width=\columnwidth]{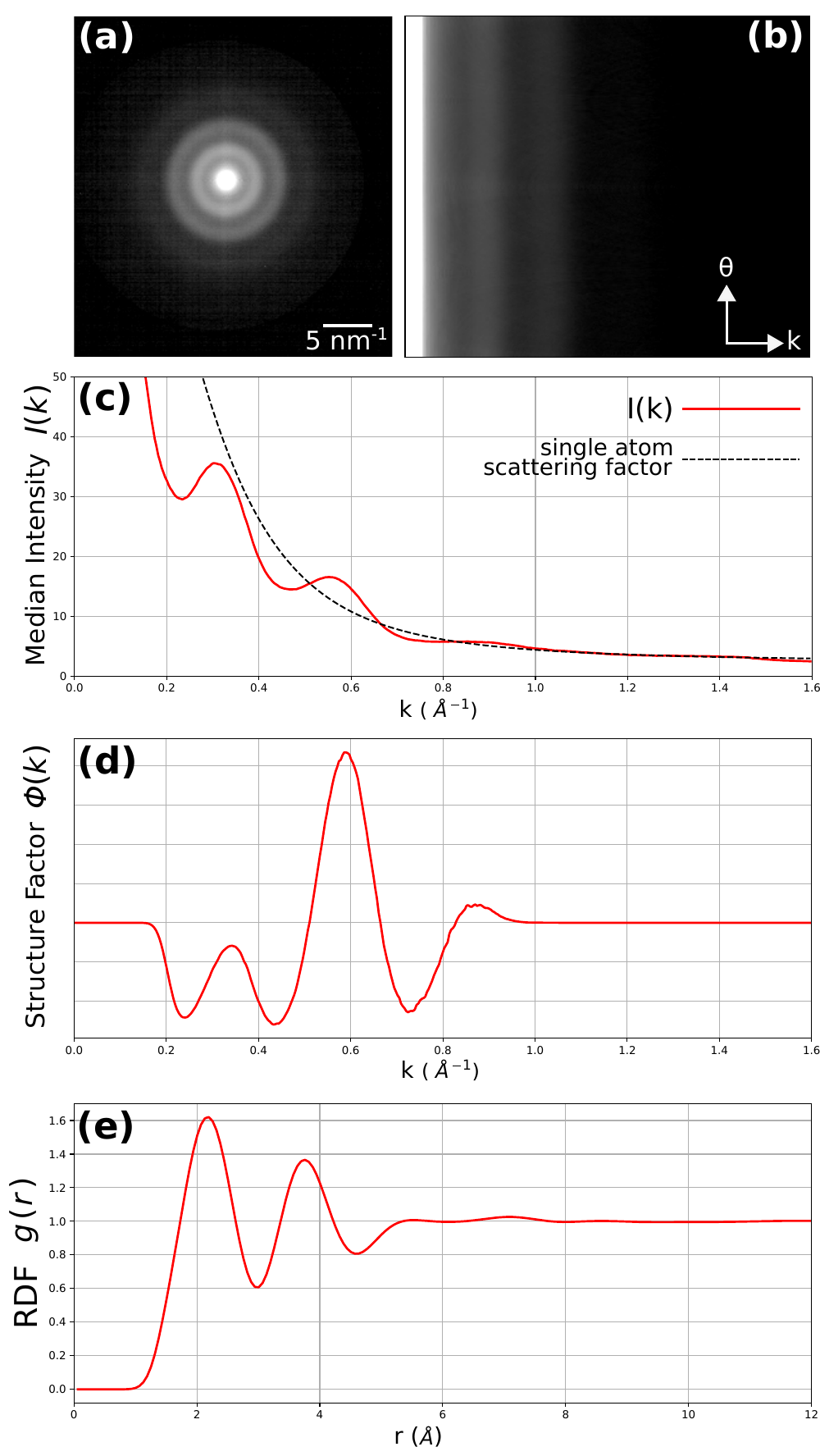}
    \caption{Radial distribution function of amorphous silicon.  (a) An average diffraction pattern.  (b) The polar-elliptical transform of (a).  (c) The radial intensity profile, calculated from (b).  (d) The structure factor, calculated by determining and subtracting off the single atom scattering factor, and applying a bandpass mask to cut off high and low frequency noise.  (e) The radial distribution function, calculated using the structure factor, showing the first few shells of Si atoms.}
    \label{F:RDF}
\end{figure}

The RDF can be directly determined from the average diffraction pattern of an amorphous material, as long as enough counts / images are collected to average out any local density fluctuations, and the probe convergence semiangle is sufficiently small to not blur out the diffraction pattern \cite{egami2003underneath}. 
An example of the mean diffraction pattern from amorphous silicon is shown in Fig.~\ref{F:RDF}a.
A radial integral is then calculated, here using a polar-elliptical methods of Sec.~\ref{S:polar_transformation}, yielding the diffracted intensity as a function of distance from the optic axis.
The resulting curve, $I(k)$, is shown in Fig.~\ref{F:RDF}c.
The important elements of this signal are (1) thermal diffuse background, resulting from thermal motion of the atoms and which dominates the behavior shown here at low $k$ values, (2) the single atom scattering factors, describing the scattered intensity profiles which result from individual atoms and which dominate the behavior at high $k$ values, and (3) the structure factor $\Phi(k)$, which describes the arrangement of atoms relative to one another in the material.
By fitting the thermal background and atomic scattering factors it is possible to calculate the structure factor, and from the structure factor it is possible to calculate the RDF.
Figure~\ref{F:RDF}d and e show the structure factor and RDF, respectively.

We ultimately invert the structure factor, a diffraction space quantity, to retrieve the RDF, a real space quantity.
The sampling of the RDF is thus determined by the maximum $k$ values in the experimental data.
In py4DSTEM we therefore upsample by padding the structure factor with zeros before inversion, which allows extraction of an RDF which is in principle arbitrarily smooth.
However, that smoothness should not be over-interpreted: the highest frequencies at which true information has been transferred is set by the maximum $k$ from the experimental data.

In gathering data for RDF analysis it is important to capture high scattering angles to use in fitting the atomic scattering factors, therefore fairly short camera lengths are recommended.

\subsection{Fluctuation Electron Microscopy}
\label{S:FEM}

\begin{figure}
    \centering
    \includegraphics[width=0.8\columnwidth]{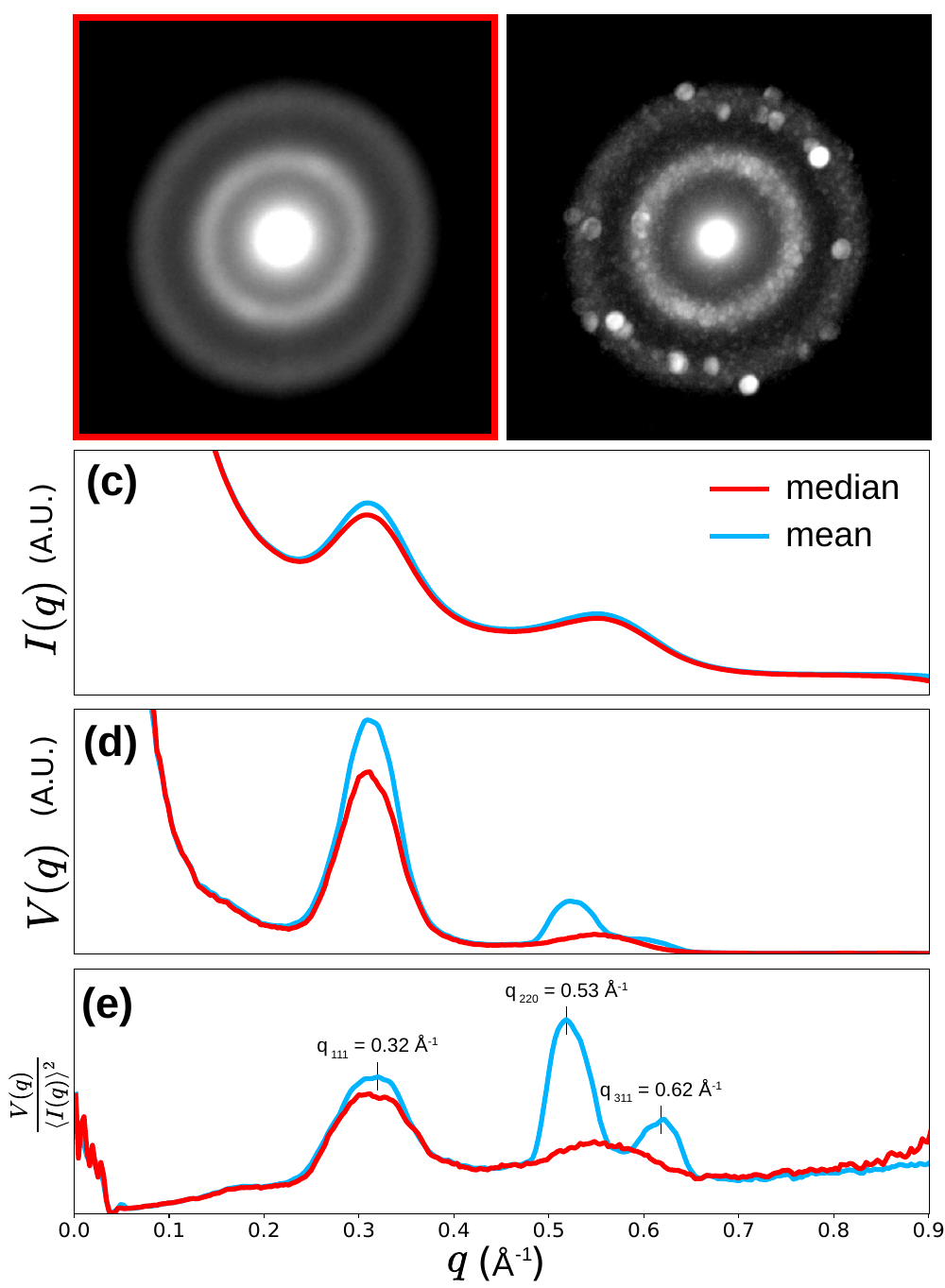}
    \caption{Fluctuation electron microscopy of predominantly amorphous silicon.  (a) An average diffraction pattern of the 4D-STEM scan, using median statistics.  (b) A diffraction image generated by selecting the maximum value at each pixel across all scan positions, and revealing the presence of some amount of Bragg scattering, and therefore crystallinity, in this sample.  (c) The radial intensity profiles of the dataset using median (\textit{red}) and mean (\textit{blue}) statistics respectively.  (d) The variance over all angles versus radial position.  (e) $\frac{V(q)}{\langle I(q) \rangle^2}$, a measure of short and medium range order which becomes larger with increasing order.  Note the two peaks in blue (mean statistics) which become a single broad peak in red (median statistics). These come from Bragg scattering, suggesting the median statistics are superior for evaluating the amorphous structure.}
    \label{Fig:FEM}
\end{figure}

Fluctuation electron microscopy (FEM) is a method which, like RDF analysis, is used to characterize the structure of amorphous materials.
In RDF analysis structure is typically considered out to distances of perhaps the first few shells of neighboring atoms, considered the ``short range order'' regime.
However, many amorphous materials have a substantial degree of structural ordering beyond the first few shells \cite{phillips1979topology}.
This property is referred to as ``medium range order'' in materials science \cite{treacy1996variable, gibson2000atom}.
When using 4D-STEM to study amorphous materials, the STEM probe size (set by the convergence semiangle and/or probe defocus) can be tuned to match the size of atomic clusters.
When these clusters deviate from a fully random distribution, they lead to ``speckles'' in the amorphous halo.
The technique of quantifying the degree of variability as a function of scattering angle and probe size is called fluctuation electron microscopy (FEM) \cite{voyles2002fluctuation}. 
In this section we qualitatively discuss FEM, and a mathematical treatment is in Appendix~\ref{A:FEM}.

Our approach follows the methods of \cite{bogle2010size}.
The idea is to calculate the variance $V(k)$ of the diffraction patterns as a function of scattering angle.
With an appropriate normalization (see the appendix), the variance can be thought of as a metric of order.
Consider the limiting cases: in a minimally ordered sample the atomic distribution is completely homogeneous, leading to perfectly smooth diffracted rings and thus zero variance at a given scattering angle.
In a maximally ordered sample, a perfect crystal, the rings resolve into Bragg disks, so that the variance at some fixed $k$ containing peaks will be maximized.
The RDF is primarily sensitive to the 2-body atomic pair correlations, whereas the FEM variance is more sensitive to 4-body pair-pair correlations \cite{treacy1996variable, rodenburg1999measurement}, hence its utility in examining medium range order.


Figure~\ref{Fig:FEM} shows an FEM measurement of an amorphous silicon sample, performed in py4DSTEM.
Figure~\ref{Fig:FEM}a shows the mean diffraction pattern of the dataset, with two strong amorphous rings visible.
However, plotting the maximum intensity across all probe positions as in Fig.~\ref{Fig:FEM}b shows some Bragg disk features, due to small regions of crystallinity in some probe positions.
We could simply exclude these patterns from the FEM measurement, but there is a simpler way to suppress or eliminate unwanted contributions of crystalline regions to the variance $V(q)$: replace the mean intensity as a function of orientation angle $\phi$ with the median intensity.
Median statistics are much more robust against outliers, such as the high variance due to Bragg peaks.

Figure~\ref{Fig:FEM}c-e show FEM measurements using both mean and median statistics.
The presence of crystalline regions barely effects the mean intensity $\langle I(q) \rangle_{\phi,\mathbf{R}}$, but strongly modulates the variance $V(q)$.
Figure~\ref{Fig:FEM}e shows a strong signal at $q = 0.318$ \AA$^{-1}$ that corresponds to the distribution of nearest neighbor atoms in amorphous Si, with a mean scattering vector approximately equal to the crystalline Si [111] lattice spacing.
This signal is approximately the same using both mean and median statistics, unlike the second peak.
When the normalized variance is calculated using mean statistics, two additional Bragg peak signals become visible on top of the second broad amorphous peak, corresponding to the [220] and [311] crystalline Si diffraction peaks.
This example highlights the importance of either careful inspection of the diffraction images or using robust statistical methods such as medians when performing FEM studies.


\subsection{Phase retrieval}
\label{S:phase_retrieval}

py4DSTEM includes two methods for reconstruction of the sample potential:
differential phase contrast (DPC) and ptychography.
Broadly, the idea in both methods is to extract the amount of extra phase that has been added to the electron beam at each scan position.
That phase is then taken to be the total (i.e. projected) sample potential at this scan position times a constant which encodes electron-charge interaction strength.
Figure~\ref{F:dpc} show the results of the py4DSTEM phase retrieval algorithms applied to a carbon nanotube \cite{yang2016simultaneous}.

These methods make the transmission function approximation and the projection approximation, and thus can be expected to be most reliable for thin, low-$Z$ samples.
In cases where these conditions don't hold, phase retrieval should be interpreted cautiously.

\subsubsection{Differential phase contrast}
\label{S:DPC}

\begin{figure*}
    \centering
    \includegraphics[width=\textwidth]{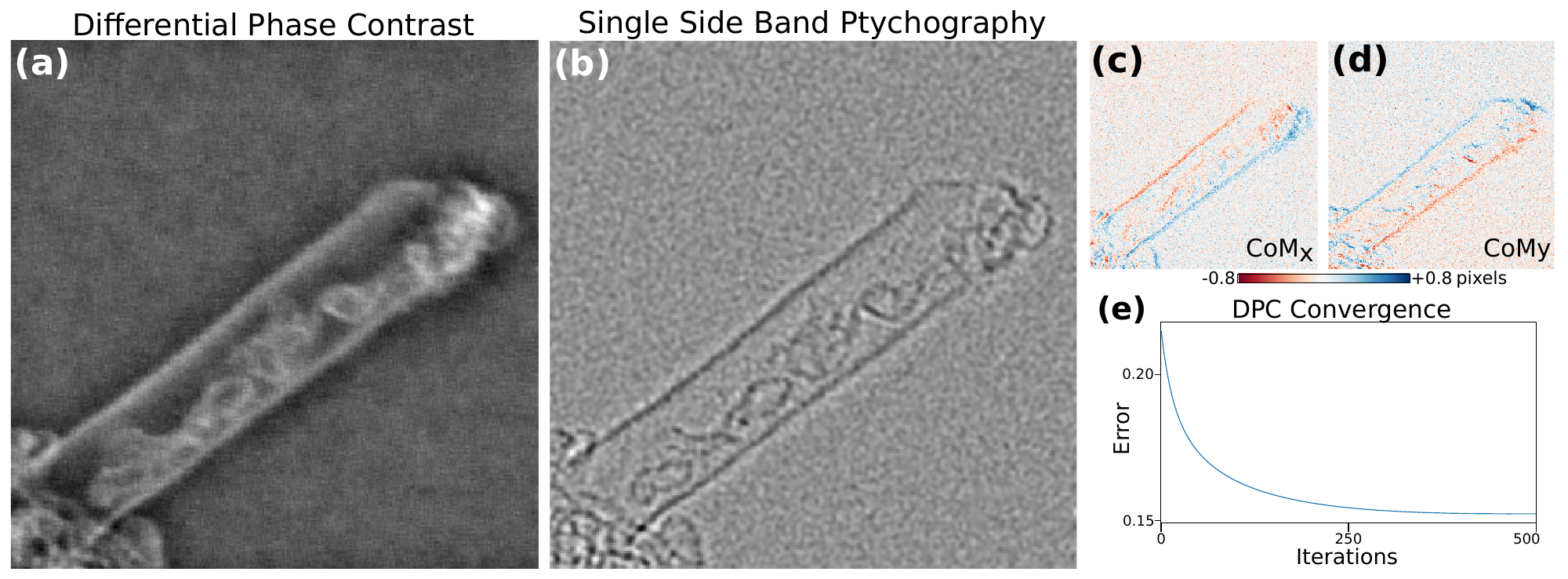}
    \caption{Phase retrieval of a carbon nanotube, using (a) differential phase contrast, or (b) single side band ptychography.  (c,d) The centers of mass of the diffraction patterns, used in DPC.  (e) Convergence of the boundary condition correction component of the DPC algorithm.}
    \label{F:dpc}
\end{figure*}

Differential phase contrast (DPC) uses the fact that, for a sufficiently thin object, the mean deflection of the electron probe at each scan position in the STEM raster is related to the specimen electric and magnetic field components transverse to the beam propagation direction.
Examples of material science applications of this technique include the study of built-in electric fields in semiconductor devices~\cite{shibata2015imaging}, magnetic skyrmions~\cite{matsumoto2016direct} and domain structures~\cite{chapman1978direct} and as a technique for efficient visualisation of light atoms in materials~\cite{song2019atomic}.
The technique was first suggested by Dekkers and De Lang~\cite{dekkers1974differential}, was extensively applied to the study of magnetic materials from the 1970s onward by Chapman and colleagues and has seen more ubiquitous use with increased uptake of more sophisticated segmented detectors~\cite{shibata2012differential} and the advent of fast-readout electron cameras in STEM~\cite{krajnak2016pixelated}. %

Figure~\ref{F:dpc}a shows the sample potential after DPC reconstruction.
The mean probe deflections are shown in Fig.~\ref{F:dpc}c,d at each scan position in the $x$- and $y$-directions, respectively.
Once these are calculated, optionally after defining some mask to cut off high angle scattering, DPC considers this vector field of deflections to be the gradient of some scalar function.
The primary task of DPC is thus to reconstruct the scalar field (a.k.a. the DPC image) which has as its gradient the measured probe deflections.
In py4DSTEM this inversion is accomplished by Fourier integration of the probe deflections~\cite{arnison2004linear,close2015towards}.
In the phase object regime (also known as the multiplicative approximation), the resulting scalar field is proportional to the sample potential.
Appendix~\ref{A:dpc} derives the relation between the beam deflections and the sample potential, and discusses the Fourier integration approach used.
A consequence of Fourier integration is that it implicitly assumes periodic boundary conditions, which can be problematic for non-periodic electron microscopy specimens. Boundary condition handling is important to a high quality DPC reconstruction, and in py4DSTEM we have used an iterative boundary condition correction algorithm which is discussed in detail in Appendix~\ref{A:dpc}.

The rotational offset between real and diffraction space needs to be correctly calibrated to perform the Fourier integration step.
One possibility is to use the method discussed in Sec.~\ref{S:calibration}.
In the context of DPC, alternative approaches to this calibration are also possible.
In one, the beam deflections are assumed to be a conservative vector field, which must be true if they are the gradient of a scalar field.
However, if the coordinate systems of real and reciprocal space contain a relative rotation, the measured beam deflections will all be similarly rotated, resulting in general in a non-conservative field.
The correct rotational offset can thus be identified by finding the relative rotation which results in beam deflections which are conservative, which can be identified by minimizing the curl as a function of the rotation.
In another approach, we note that the contrast of a DPC reconstruction, that is the contrast of the scalar field which results from Fourier integration of the beam deflections, is typically maximized when the rotational offset is correct.
Thus the calibration may also be performed by maximizing the DPC contrast as a function of real/diffraction space rotation.
Note that this method permits a 180 degree ambiguity in the rotational offset, corresponding to a contrast reversal in the DPC image.

Finally, we note that Fourier integration effectively applies a low pass filter (see Appendix~\ref{A:dpc}).
Some amount of low pass filtering is therefore inherent in DPC imaging as implemented in py4DSTEM.

\subsubsection{Ptychography}
\label{S:ptychography}

Phase retrieval is difficult because phase is never directly recorded; instead, the detector only captures the square modulus of the electron wavefunction.
In electron ptychography of crystals, the idea is that with a large enough convergence angle, the central disk will begin to overlap with other Bragg disks.
In the overlap regions the phases of the two beams add coherently, and consequently phase reconstruction is possible by analyzing these regions.
The method is analogous to holography, which combines a scattered beam and a reference beam to create an interference pattern, except that the `scattered' and `reference' beams are now the central beam and the Bragg reflected beams.
Variations in these regions of interference as the beam is scanned enable phase retrieval.
Ptychography was first suggested as a method to solve the crystallographic phase problem by Hoppe
\cite{hoppe1969diffraction1,hoppe1969diffraction2,hoppe1969diffraction3}, and later extended to solve the phase problem for arbitrary specimens by Rodenburg \cite{rodenburg1992theory}.

py4DSTEM includes a ptychographic reconstruction algorithm which calculates the phase in a single step, based on the single-side-band approach and discussed in detail in Appendix~\ref{A:ptychography}. 
Figure~\ref{F:dpc}b shows the results for the carbon nanotube discussed above, and clearly reveals both the walls of the tube as well as the tortuous structure of carbon inside the tube.
In general, direct solvers tend to be fast, however, better reconstruction quality is usually achieved with iterative algorithms.
Unfortunately, a patent on iteration creates substantial challenges to making iterative ptychography codes of any sort freely available to the scientific community.


\section{Conclusion}
\label{S:conclusions}

In this paper, we have presented the py4DSTEM software package written in Python, for analysis of 4D-STEM experiments.
We have described the program's purpose and structure, including an HDF5 based file standardization for 4D-STEM.
We've described how py4DSTEM can be used for preprocessing and calibrating data, finding Bragg disk positions, transformation into polar-elliptical coordinates, and for classifying diffraction patterns based on commonalities in their diffraction patterns.
We demonstrated measurements including virtual imagining, phase mapping, mapping strain in crystalline and amorphous materials, RDF and FEM analyses, and phase reconstruction with DPC and with ptychography.
The analysis here spans 8 datasets, including seven experimental and one synthetic dataset.

The py4DSTEM codes and many examples are freely available in the \href{https://github.com/py4dstem/py4DSTEM}{py4DSTEM repository} on Github.
As an open source project, both new users and new contributors are enthusiastically encouraged to try the code, use it in your own work, or make a pull request.

\section*{Acknowledgements}

This project is supported by the Toyota Research Institute.
Work at the Molecular Foundry was supported by the Office of Science, Office of Basic Energy Sciences, of the U.S. Department of Energy under Contract No. DE-AC02-05CH11231.
BHS and LAH were supported by the Toyota Research Institute.
SEZ was supported by STROBE, an NSF Science and Technology Center, under Grant No. DMR 1548924. 
SZ was supported by the US Office of Naval Research under Grant No. N00014-17-1-2283.  
JD was supported by the Dow University Partnership Initiative Program.
CO and JC acknowledge support of Department of Energy Early Career Research Awards.
The authors thank Shreyas Cholia, Matthew Henderson, Rollin Thomas, and Ludovico Bianchi from the National Energy Research Scientific Computing Center (NERSC) for support with high-performance computing integration.
NERSC is a U.S. Department of Energy Office of Science User Facility operated under Contract No. DE-AC02-05CH11231.
The authors thank Blas Uberuaga for sample acquisition and support.
Thanks also to the developers of hyperspy, openNCEM, scipy, numpy, and matplotlib, without whom this project would not have been possible.

\appendix
\section*{Appendices}

\section{Basic formalism for 4D-STEM}
\label{A:STEM}

\begin{figure}
    \centering
    \includegraphics[width=1\columnwidth]{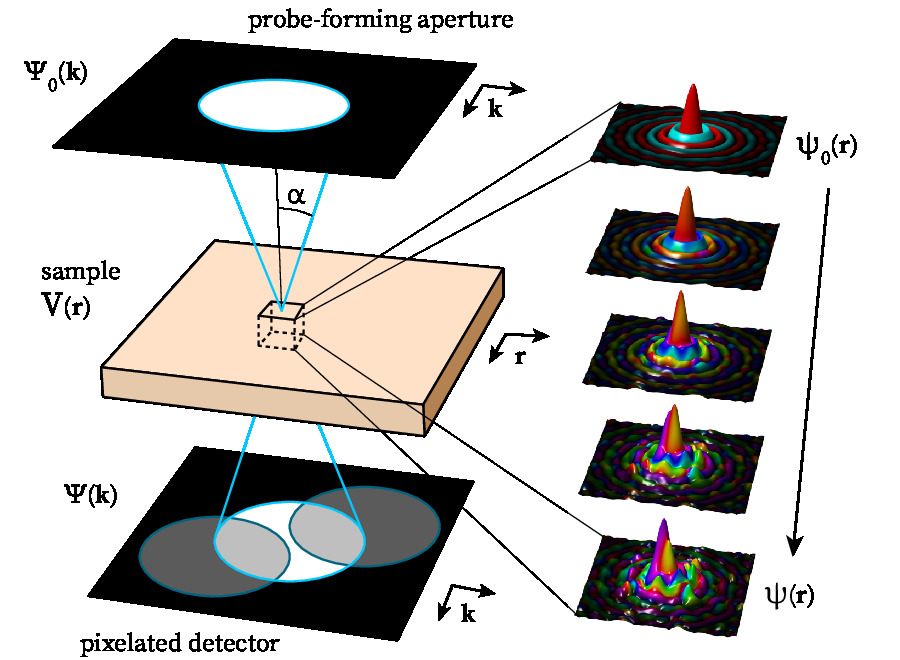}
    \caption{Schematic of STEM experimental geometry, showing initial probe focused onto sample surface, propagating through sample, exiting the sample, and finally being imaged in the far field detector plane.}
    \label{F:stem_geometry}
\end{figure}

Figure~\ref{F:stem_geometry} shows the geometry of a STEM experiment.
We define the following probe wavefunctions:
\begin{align*}
    \Psi_\text{0} (\mathbf{k}) &= \text{initial probe formed in diffraction space.}  
    \\
    \psi_\text{0} (\mathbf{r}) &= \text{probe focused onto sample surface.}
    \\
    \psi(\mathbf{r}) &= \text{probe at exit plane of sample.}  
    \\
    \Psi (\mathbf{k}) &= \text{far field probe in detector plane.}
\end{align*}
where $\mathbf{r}$ and $\mathbf{k}$ denote coordinates in the real space image and diffraction space planes, respectively, and the diffraction plane coordinate $|\mathbf{k}| = \frac{\alpha}{\lambda}$ for scattering angle $\alpha$ and relativistically corrected electron wavelength $\lambda$.
In a STEM experiment, scan coils are used to move the probe to a given position $\mathbf{R}$, denoted by $\psi_0 (\mathbf{r} - \mathbf{R})$.
In this appendix, we first describe a simple model for 4D-STEM datasets, which primarily refers to the diffraction plane wavefunction $\Psi(\mathbf{R},\mathbf{k})$.
We then briefly discuss the more general question of how the wavefunction evolves from the initial probe $\Psi_0$ to the final probe $\Psi$ on the detector.

A 4D-STEM dataset typically takes the form of a four-dimensional array of intensity values,
\begin{align*}
    I &= I_{i,j,n,m} \\
      &= I(R_x,R_y,k_x,k_y) \\
      &= I(\mathbf{R},\mathbf{k}).
\end{align*}
Here, each $I_{i,j,n,m}$ is a scalar and $(i,j,n,m) \in \mathbb{N}$, i.e. the dataset is a discrete 4D grid of numbers.
The correspondences between $(i,j)$ and scan position $\mathbf{R} = (R_x,R_y)$ and between $(n,m)$ and diffraction coordinate $\mathbf{k} = (k_x,k_y)$ are determined by the real and diffraction space pixel size calibrations.
The value of each $I_{i,j,n,m}$ is given by the electron flux passing through the appropriate detector pixel, or by the square modulus of the beam wavefunction integrated over the detector pixel at $\mathbf{k}$ when the beam raster position is $\mathbf{R}$.
Thus the 4D-STEM dataset may be modelled by
\begin{align*}
\label{E:4D-STEM}
    I(\mathbf{R},\mathbf{k}) &=
    \int_{k_x}^{k_x+\Delta k}\int_{k_y}^{k_y+\Delta k} \lvert \Psi(\mathbf{R},\mathbf{k})\rvert^2 dk_xdk_y \\
         &\approx \lvert \Psi(\mathbf{R},\mathbf{k})\rvert^2
\end{align*}
where the approximation is exact in the limit of infinitesimally small detector pixels.
Note that this simple model does not account for finite information transfer in the microscope, which could be included with a multiplicative transfer function $M(\mathbf{k})$.

In a STEM experiment with an integrating detector (ADF, BF, etc.), the image $I(\mathbf{R})$ can be modeled as
\begin{equation}
\label{E:stem_imaging}
    I(\mathbf{R}) = \int 
    | \Psi(\mathbf{R},\mathbf{k}) |^2 D(\mathbf{k}) d\mathbf{k}
\end{equation}
where $D(\mathbf{k})$ reflects the detector geometry.
For some 4D-STEM signal $I(\mathbf{R},\mathbf{k})$ we can write down an equivalent virtual image $I_\text{v}(\mathbf{R})$ as: \begin{equation}
\label{E:virtual_imaging}
    I_\text{v}(\mathbf{R}) 
    = \int I(\mathbf{k},\mathbf{R}) D(\mathbf{k}) d \mathbf{k}.
\end{equation}
If Eqs.~\ref{E:stem_imaging} and \ref{E:virtual_imaging} look similar, it's because they are.
The key difference is in the meaning of the integration over $D(\mathbf{k})$: in the former equation, it describes the action of the detector, and the integration occurs in hardware during data acquisition; in the latter equation, it is a prescription for which pixels of the 4D datacube need to be summed in post-processing.

The evolution of the probe is comparatively simple from the probe forming aperture to the sample plane, and from the sample plane to the detector - both are given by Fourier transforms:
\begin{align}
\label{E:diffraction}
    \psi_0 (\mathbf{r}) 
    &=\mathcal{F}_{\mathbf{k}\to\mathbf{r}}
    \Psi_0 (\mathbf{k}) 
    \nonumber \\
    \Psi (\mathbf{k}) 
    &=\mathcal{F}_{\mathbf{r}\to\mathbf{k}}
    \psi (\mathbf{r}) 
\end{align}
where $\mathcal{F}_{\mathbf{r}\to\mathbf{k}}$ is the forward transform from the real to diffraction domain, and $\mathcal{F}_{\mathbf{k}\to\mathbf{r}}$ is the inverse transform from the diffraction to real domain.
The most common initial condition for the electron probe in 4D-STEM is given by a circular aperture in a condenser plane
\begin{eqnarray}
    \Psi_0 (\mathbf{k}) &=& A(k_\text{max}) 
    \nonumber \\
    &=& 
    \left\{ \begin{array}{rl}
    1 &\mbox{ if $|\mathbf{k}| \leq k_\text{max}$} \\
    0 &\mbox{ otherwise}
       \end{array} \right.
\end{eqnarray}
where $A(k_\text{max})$ is the 2D ``top hat'' function, and $k_\text{max}$ is the maximum scattering vector of the probe.
The probe incident on the sampe is then an Airy disk function
\begin{equation}
    \psi_0 (\mathbf{r}) = \frac{\text{J}_1 
    (2 \pi k_\text{max} |\mathbf{r}|)}{\sqrt{\pi} |\mathbf{r}|}
\end{equation}
where $\text{J}_1$ is a Bessel function of the first kind, and the peak amplitude is equal to $\sqrt{\pi} k_\text{max}$.
This function is shown graphically in the upper right corner of Figure~\ref{F:stem_geometry}.
More complex STEM probes can be formed by using amplitude-patterned apertures \cite{guzzinati2019electron, zeltmann2019patterned}, phase plates \cite{mcmorran2011electron, ophus2016efficient, yang2016enhanced, verbeeck2018demonstration}, or other methods \cite{blackburn2014vortex, pozzi2017generation}.
In a vacuum, $\psi_0(\mathbf{r}) =  \psi(\mathbf{r})$, so that $\Psi(\mathbf{k})$ and $\Psi_0(\mathbf{k})$ are identical up to scaling and phase factors, so that without a sample the image on the detector directly reflects the electron beam passing through the probe-forming aperture.

The change in the wavefunction from $\psi_0(\mathbf{r})$ to $\psi(\mathbf{r})$, as the beam passes through a sample is, in general, analytically intractable, so numerical methods are typically used.
Using some approximations described in \cite{kirkland2010advanced}, and omitting the scan coordinate $\mathbf{R}$ for clarity, the interaction of the STEM probe with the sample is governed by the time-independent Schr\"{o}dinger equation
\begin{equation}
\label{E:schrodinger}
    \frac{\partial \psi(\mathbf{r})}{\partial z} 
    = 
    \frac{\ii \lambda}{4 \pi} 
    \left[ 
    \frac{\partial^2 \psi(\mathbf{r})}{\partial x^2}  + 
    \frac{\partial^2 \psi(\mathbf{r})}{\partial y^2} 
    \right]
    +
    \ii \sigma V(\mathbf{r}) \psi(\mathbf{r})
\end{equation}
where $\ii$ is the imaginary constant, $\sigma$ is the relativistically-corrected electron-matter interaction constant, and $V(\mathbf{r})$ is the electrostatic potential inside the sample.
Because the two operators on the right-hand side of equation \ref{E:schrodinger} do not commute, it is typical to use a split-step method to numerically solve this equation called the multislice method, first derived in \cite{cowley1957scattering}.
To use the multislice method to solve the interaction of the electron beam with the sample, we first divide up the sample into a series of $N$ slices, $V_n(\mathbf{r})$, which are 2D arrays that integrate all of the electrostatic potential contained in a given slice of thickness $\Delta z$, given by
\begin{equation}
     V_n(\mathbf{r}) =
     \int_{z - \Delta z/2}^{z + \Delta z/2}
     V(\mathbf{r}) dz
\end{equation}
By assuming each slice has infinitesimal thickness, the solution to the transmission operator is given by
\begin{equation}
    \label{E:multiplicativeApprox}
    \psi(\mathbf{r})
    = T(\mathbf{r}) \psi_0(\mathbf{r}) =
    e^{\ii \sigma V_n(\mathbf{r})} \psi_0(\mathbf{r}).
\end{equation}
Between each slice, we assume zero electrostatic potential and can therefore advance the electron wave by using the free-space propagation operator, which can be efficiently applied in Fourier space \cite{kirkland2010advanced}
\begin{equation}
    \psi(\mathbf{r})
    =
    \mathcal{F}_{\mathbf{k}\to\mathbf{r}}
    \left\{
    e^{\ii \lambda \Delta z |\mathbf{k}|^2} 
    \Big[
    \mathcal{F}_{\mathbf{r}\to\mathbf{k}} \psi_0(\mathbf{r}) 
    \Big] \right\}.
\end{equation}
Note that the propagation operator $e^{\ii \lambda \Delta z |\mathbf{k}|^2}$ uses the 2D inverse spatial coordinate $\mathbf{k} = {k_x}^2 + {k_y}^2$.
We alternate the application of the transmission and propagation operators to calculate the final wavefunction after interacting with the sample,  
\begin{equation}
    \psi(\mathbf{r}) 
    = \left[
    \prod_{n=1}^N 
    \Big\{
    \mathcal{F}_{\mathbf{k}\to\mathbf{r}}
    \Big[
    e^{\ii \lambda \Delta z |\mathbf{k}|^2} 
    \Big\{
    \mathcal{F}_{\mathbf{r}\to\mathbf{k}}
    \Big[
    e^{\ii \sigma V_n(\mathbf{r})}
    \Big] \Big\} \Big] \Big\} \right]
    \psi_0(\mathbf{r}), \nonumber
\end{equation}
which is typically referred to as the exit wave.
This method requires $N$ transmission operations and $N-1$ propagation operations.
The multislice method is often used for modeling 4D-STEM experiments, but can require a prohibitively high amount of computation time for very large simulations.
Recently, a more efficient method has been developed to simulate 4D-STEM experiments called PRISM \cite{ophus2017fastSim}, which has been made available as a simulation code \cite{pryor2017streaming}, and extended to simulate electron energy loss spectroscopy (STEM-EELS) inelastic scattering as well \cite{brown2019linear}.

\section{Cross, phase, and hybrid correlations}
\label{A:cc}

Cross-correlative template matching is a standard tool in image processing, and is widely used in computational analysis for electron microscopy \cite{frank1975averaging, modersitzki2004numerical}.
The purpose of this appendix is to outline the formalism for these methods, and to briefly discuss the effects of and appropriate uses cases for so-called `hybrid' correlations.

For functions $f$ and $g$, written in one dimension for simplicity, the cross correlation is defined as
\begin{equation}
\label{E:cc1}
    (f \star g)(x) = \int_\infty^\infty f(y)^*g(x+y)dy
\end{equation}
where $^*$ indicates complex conjugation.
The key idea here is that if $f(x-a) = g(x)$, then $(f \star g)(x=a)$ will be a maximum, because the integrand then becomes $\lvert f(y)\rvert^2$ and two functions are perfectly overlapped.
Therefore, the cross correlation of the vacuum probe template with a diffraction pattern can be used to extract the Bragg disk positions simply by identifying the cross correlation maxima.

Computationally, this is implemented via the cross correlation theorem, which states that
\begin{equation}
\label{E:cc2}
    (f \star g)(x) = \mathcal{F}^{-1}\left( (\mathcal{F}f)^* (\mathcal{F}g) \right)
\end{equation}
where $\mathcal{F}$ is the Fourier transform.
This follows directly from the Fourier transform of Eq.~\ref{E:cc1} and the change of variables $x'=x+y$.
Equation~\ref{E:cc2} therefore allows the integral of Eq.~\ref{E:cc1} to be computed efficiently via a few FFT operations, which is important because performing the cross correlation on each diffraction pattern (often 10,000 or more) is the most computationally intensive step of many analysis workflows.

In contrast to Eq.~\ref{E:cc2}, the so-called phase correlation normalizes by the amplitude in Fourier space before applying the inverse transform:
\begin{equation}
\label{E:cc3}
    (f \star g)_\text{phase}(x) = \mathcal{F}^{-1}\left(   \frac{(\mathcal{F}f)^* (\mathcal{F}g)}  {\lvert(\mathcal{F}f)^* (\mathcal{F}g)\rvert}   \right)
\end{equation}
This leads to an analytically pleasing result: now, if $f(x-a) = g(x)$, then $(f \star g)_\text{phase}(x) = \delta(x-a)$.
The result follows directly from substituting $f(x-a) = g(x)$ in to Eq.~\ref{E:cc3} and making use of the Fourier shift theorem.
Thus where the cross correlation simply has a maximum where $f$ and $g$ best overlap, the phase correlation yields a delta function which selects the point of interest.
As a practical matter, however, phase correlations are also highly sensitive to noise, and this application tend to lead to many false positives when used with real data.

\begin{figure*}
    \centering
    \includegraphics[width=\textwidth]{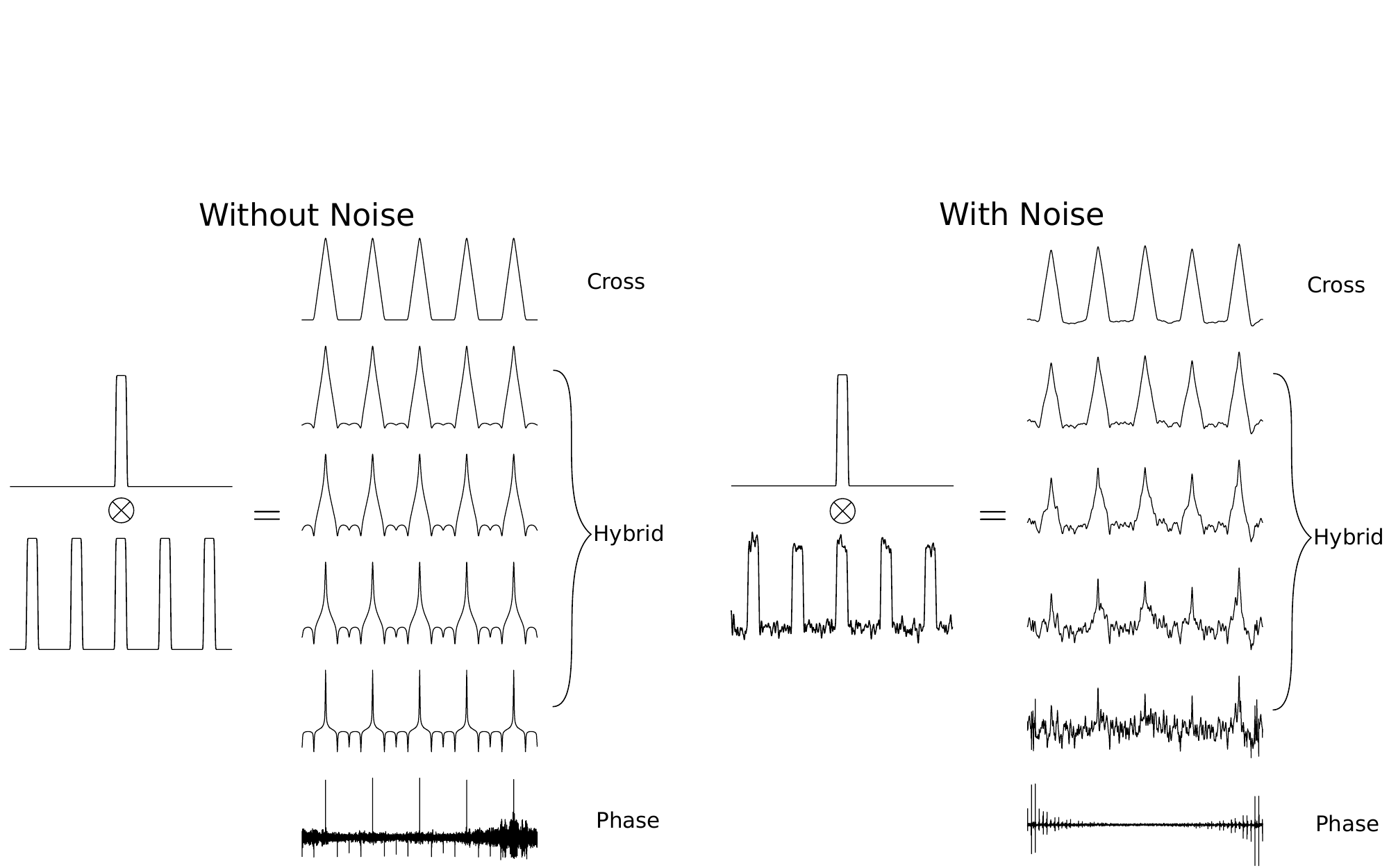}
    \caption{Cross correlation, phase correlation, and hybrid correlation.  A 1D vacuum probe has been correlated with a 1D comb of Bragg disks without (left) and with (right) noise.  Various values of $n$ are used, where $n=1$ is a cross correlation, $n=0$ is a phase correlation, and intermediate values are `hybrid' correlations -- see Eq.~\ref{E:cc4}.  Hybrid correlations increase sensitivity to edges and narrow the central maxima relative to a cross correlation, but also increase sensitivity to noise.}
    \label{F:cc}
\end{figure*}

An intermediate approach is possible using a hybrid correlation, in which a normalization somewhere in between a phase and cross correlation is used, as follows:
\begin{equation}
\label{E:cc4}
    (f \star g)_\text{n}(x) = \mathcal{F}^{-1}\left(   \frac{(\mathcal{F}f)^* (\mathcal{F}g)}  {\lvert(\mathcal{F}f)^* (\mathcal{F}g)\rvert^{1-n}}   \right)
\end{equation}
Here, $n \in [0,1]$.
For $n=1$, the result is a cross correlation, and for $n=0$, the result is a phase correlation.
Intermediate values may be thought of is applying intermediate weighting to the amplitude versus phase components of the signals in Fourier space.
While the hybrid correlation is a heuristic approach, it is often effective.
Giving more weight to the phase components (lower values of $n$) increases sensitivity to edges and can do a better job of identifying faint Bragg disks, however, the trade off is typically an increase in false-positives.
Experience with many datasets indicates that an $n$ value in the neighborhood 0.85 or 0.9 -- similar to a cross correlation, but with a slightly increased sensitivity to edges -- frequently yields good results. 
The noisier the data, the more caution is in order in using lower $n$ values, and for very noisy data pure cross-correlations are recommended.
Figure~\ref{F:cc} shows cross, hybrid (for several $n$ values) and phase correlations in one dimension for simulated data with and without noise.

\section{Bragg vector map formalism}
\label{A:bvm}

Let us refer to the $i$'th Bragg disk detected in the diffraction pattern at scan position $(R_x,R_y)$ as $\mathcal{B}_{R_x,R_y,i}$.
In computer memory, this might be thought of as a length 3 tuple:  $\mathcal{B}_{R_x,R_y,i} = (k_{x,i},k_{y,i},I_i)$, where the subscript $i$ indicates the $i$'th peak, and the three values are the coordinates of the disk center in diffraction space and the disk's intensity.
Analytically, we can think of the $\mathcal{B}_{R_x,R_y,i}$ as Kronecker deltas of strength $I_i$:
\begin{equation}
\label{E:braggdisks}
    \mathcal{B}_{R_x,R_y,i}(\mathbf{k}) = I_i\delta(k_x - k_{x,i})\delta(k_y - k_{y,i})
\end{equation}
The delta function specifies where the Bragg condition is met for parallel illumination; Bragg disks are formed by translating each point in the central disk by this vector, and may be thought of as the convolution of the aperture function with Eq.~\ref{E:braggdisks}

Let's denote the set of all Bragg disks detected at a scan position $(R_x,R_y)$ as $\mathcal{B}_{R_x,R_y}$.
For $N$ disks in $\mathcal{B}_{R_x,R_y}$ we can write
\begin{equation}
\label{E:braggdisks2}
    \mathcal{B}_{R_x,R_y}(\mathbf{k}) = \sum_{i=1}^N I_i\delta(k_x - k_{x,i})\delta(k_y - k_{y,i})
\end{equation}
Taking a summation over all scan positions gives
\begin{align}
\label{E:braggmap}
    \mathcal{B}(\mathbf{k}) &= \sum_{r_x \in R_x}\sum_{r_y \in R_y} \mathcal{B}_{r_x,r_y}(\mathbf{k})  \\
                            &=  \sum_{r_x,r_y,i} I_{r_x,r_y,i}\delta(\mathbf{k} - \mathbf{k}_{r_x,r_y,i})
\end{align}
$\mathcal{B}$ is the Bragg vector map.
Physically, is interpretable as a (unnormalized) distribution of measured Bragg vector directions found within the sample over the area of the 4D-STEM scan.

\section{Elliptical fitting and transforms}
\label{A:ellipses}

In this appendix we describe various elements of py4DSTEM that make use of or relate to elliptical coordinates.
First we discuss elliptical fitting, which is important for correction of elliptical distortions.
We then briefly describe and relate the two elliptical parametrizations used in the code.
Finally, we describe polar-elliptical transformations and radial integration.

Two primary elliptical fitting routines are available in py4DSTEM.
The first is appropriate for data that is well-described by a 1D elliptical curve - for instance, a Bragg vector map from a sample with many randomly oriented grains will typically contain elliptical rings associated with each characteristic spacing of the material.
The second is a sum of two Gaussian functions, a simple Gaussian and a `double-sided' Gaussian, and is appropriate for fitting amorphous diffraction patterns, and is a two-dimensional fit designed to capture the first amorphous halo.

For 1D elliptical curve fitting, we first define some annular region of our 2D dataset containing pixels $(k_{x_i},k_{y_i})$, each with intensity $I_i$.
We then determine the ellipse that most closely fits this data by computing
\begin{align*}
    \underset{k_{x_0},k_{y_0},A,B,C}{\arg\min} \sum_i [&A(k_{x_i}-k_{x_0})^2 + B(k_{x_i}-k_{x_0})(k_{y_i}-k_{y_0}) + \\ 
    &C(k_{y_i}-k_{y_0})^2 - 1]I_i
\end{align*}

The double-sided Gaussian function for amorphous halo fitting is defined as
\begin{align*}
    f(k_x,k_y; &I_0,I_1,\sigma_0,\sigma_1,\sigma_2,c,R,k_{x_0},k_{y_0},B,C) = \\
    &\mathcal{N}(r; 0,\sigma_0,I_0,) + \\
    &\mathcal{N}(r; R,\sigma_1,I_1)\Theta(r-R) + \\
    &\mathcal{N}(r; R,\sigma_2,I_1)\Theta(R-r)
\end{align*}
where $(k_x,k_y)$ are the coordinates and $(I_0,I_1,\sigma_0,\sigma_1,\sigma_2,c,R,k_{x_0},k_{y_0},B,C)$ are parameters,
where $\mathcal{N}(r; R,\sigma,I)$ is a Gaussian centered at $R$ with standard deviation $\sigma$ and with maximum amplitude $I$, where $\Theta$ is the Heaviside step function, and where $r$ is the radial coordinate of an elliptical system given by $r^2 = (k_x-k_{x_0})^2 + B(k_x-k_{x_0})(k_y-k_{y_0}) + C(k_y-k_{y_0})$.
When performing a fit, as before we first define an annular region in the dataset to fit, typically about the first amorphous halo.
The first term is meant to fit the decaying background, while the second and third terms fit the amorphous halo, while allowing for an asymmetrical shape on the inner/outer sides of the ring.

When fitting ellipses we use the parametrization
\begin{equation}
\label{E:ellipse1}
    1 = A(k_x-k_{x_0})^2 + B(k_x-k_{x_0})(k_y-k_{y_0}) + C(k_y-k_{y_0})^2
\end{equation}
for numerical stability.
However the parameters of this form are not the most easily geometrically interpretable, so for this reason we also make use of the alternate parametrization
\begin{align}
\label{E:ellipse2}
    k_x &= k_{x_0} + A'r\cos(\theta)\cos(\phi) - B'r\sin(\theta)\sin(\phi) \\
    k_y &= k_{y_0} + B'r\cos(\theta)\sin(\phi) + A'r\sin(\theta)\cos(\phi) \nonumber
\end{align}
where $(k_x,k_y)$ are cartesian coordinates, $(r,\theta)$ are polar-elliptical coordinates, and $(A',B',\phi)$ are parameters corresponding to the two semi-axis, and the tilt of the $A'$-axis with respect to the $k_x$-axis.
Equations~\ref{E:ellipse1} and \ref{E:ellipse2} are related by
\begin{align*}
    A' &= \sqrt{\frac{2}{A+C+\xi}} \\
    B' &= \sqrt{\frac{2}{A+C-\xi}} \\
    \xi &= (A-C)\sqrt{1+\left(\frac{B}{A-C}\right)^2} \\
    \phi &= \frac{1}{2}\tan^{-1}\left(\frac{B}{A-C}\right)
\end{align*}

Once the appropriate elliptical parameters are known, polar-elliptical transformations may be performed.
After specifying a range and sampling the new polar coordinates, each point $(r,\theta)$ is mapped to some $(k_x,k_y)$ position in Cartesian space, from which a bilinear interpolation is then used to compute the value at $(r,\theta)$.
In py4DSTEM, arrays returned after polar-elliptical transformation are numpy masked arrays, to ensure that coordinates beyond the frame of the raw data are correctly handled, and also facilitating masking data where necessary, e.g. from a beamstop.
Radial integrals are calculated by first computing the polar-elliptical (or polar) transformation then summing along the $\theta$-direction.

\section{Classification}
\label{A:classification}

The underlying principle of the classification scheme described in Sec.~\ref{S:classification} is the definition of a particular feature vector, which is useful because it efficiently encodes the key physical features of crystalline electron scattering -- the Bragg vectors.
It therefore massively reduces the size of the data before classification, while honing in on the most physically relevant element of the diffraction data.
For a calibrated Bragg vector map containing $N$ delta-like peaks, we associate with each such peak an integer $i\in\{0,...,N-1\}$.
For each diffraction pattern we generate a length-$N$ vector $v$.
The $i$'th element of $v$ is defined in one of two ways: (1) a Boolean value indicating whether this diffraction patterns contains this Bragg peak, or (2) a floating point value of the intensity of this Bragg peak in this diffraction pattern.

With these feature vectors in hand, we use matrix factorization methods to complete the classification.
First we construct the matrix $X$ which has the feature vectors $v$ as its columns.
For data with $R_N = R_{Nx} \times R_{Ny}$ diffraction patterns, $X$ has dimensions $N \times R_N$.
$X$ is then written as
\begin{equation}
\label{E:NMF}
    X = WH
\end{equation}
where $W$ has shape $N \times C$, $H$ has shape $C \times R_N$, and $C$ is the number of classes.
$W$ may be thought of as a collection of $C$ column vectors, each describing a class in terms of weights for the various possible Bragg vectors observed over the entire dataset.
$H$ may be thought of as a set of column vectors which describe how to obtain, through linear combination of the classes, good approximations for each of the observed diffraction patterns.
Alternatively, the row vectors of $H$ may be thought of as an image (albeit reshaped): there are $C$ of them, and each describes how much to weight the corresponding class in each of the $R_N$ positions of the electron beam.
In Secs~\ref{S:classification} and \ref{S:phase_mapping} we use an algorithm based on the frequency of co-occurance of Bragg peaks across diffraction patterns to set initial values for $W$ and $H$.
We then optimize using non-negative matrix factorization \cite{scikit-learn}.

\section{Strain}
\label{A:strain}

In this appendix we discuss how the crystalline and amorphous strains are calculated in Secs.~\ref{S:crystalline_strain_mapping} and \ref{S:amorphous_strain_mapping}.

We're interested in the infinitesimal strain matrix, where the deformed lattice differs very little from the undeformed lattice.
For a material with a deformed state characterized by some displacement field $\mathbf{u}$, and considering the system in a coordinate system with abscissa and ordinate $(x_1,x_2)$, the infinitessimal strain matrix is
\begin{equation}
\label{E:strain_matrix}
    \mathbf{\epsilon} = 
               \begin{pmatrix}
               \epsilon_{11}   &   \epsilon_{12}   \\
               \epsilon_{21}   &   \epsilon_{22}
               \end{pmatrix}
               =
               \begin{pmatrix}
               \frac{\partial u_1}{\partial x_1}   &   -\frac{1}{2}\left(\frac{\partial u_1}{\partial x_2} + \frac{\partial u_2}{\partial x_1}\right)   \\
               -\frac{1}{2}\left(\frac{\partial u_1}{\partial x_2} + \frac{\partial u_2}{\partial x_1}\right)   &   \frac{\partial u_2}{\partial x_2}
               \end{pmatrix}
\end{equation}
and is typically accompanied by $\theta_R = \frac{1}{2}\left(\frac{\partial u_1}{\partial x_2} - \frac{\partial u_2}{\partial x_1}\right)$.
The $\epsilon_{ii}$ terms represent the compressive/tensile strain along the $\widehat{\mathbf{x}}_i$ directions, with positive values corresponding to tension.
$\epsilon_{12}$ represents the shear strain, and our sign convention is chosen such that positive values correspond to the angle spanned from $\widehat{\mathbf{x}}_1$ to $\widehat{\mathbf{x}}_2$ becoming more obtuse in the deformed body.
$\theta_R$ represents the rotation of the material, with positive values corresponding to counterclockwise rotation of a right-handed coordinate system.

For both crystalline and amorphous strain, the strain matrix is calculated at each beam position by comparing two comparable measurements: one of the local structure, and one of an undeformed reference structure.
For crystalline strain, the measurement we use is a pair of reciprocal lattice basis vectors.
For amorphous strain, the measurement is a transformation of the ellipse fit to the (first) amorphous halo.


For crystalline strain, consider a real space lattice with reference basis vectors $a^0 = (\mathbf{a}^0_1,\mathbf{a}^0_2)$ and local, deformed lattice vectors $a = (\mathbf{a}_1,\mathbf{a}_2)$.
The transformation matrix $T_{a^0 \to a}$ describes the linear deformation of the space, and given the lattice vectors is calculable via
\begin{equation*}
    a = T^a a^0
\end{equation*}
The strain matrix in Eq.~\ref{E:strain_matrix} is defined with respect to some arbitrary area element of the material under study, so we consider a square unit area element with sides $(\widehat{\mathbf{e}}^0_1, \widehat{\mathbf{e}}^0_2) = (\widehat{\mathbf{x}}_1, \widehat{\mathbf{x}}_2)$.
The transformation $T^a$ maps these to a new set of vectors $(\mathbf{e}_1, \mathbf{e}_2)$.
In the limit of small area elements, the relevant derivatives are then expressible as
\begin{align*}
    \frac{\partial u_1}{\partial x_1} &= (\mathbf{e}_1 - \mathbf{e}^0_1) \cdot \widehat{\mathbf{e}}^0_1 &&= T^a_{11} - 1 \\
    \frac{\partial u_1}{\partial x_2} &= (\mathbf{e}_1 - \mathbf{e}^0_1) \cdot \widehat{\mathbf{e}}^0_2 &&= T^a_{21} \\
    \frac{\partial u_2}{\partial x_1} &= (\mathbf{e}_2 - \mathbf{e}^0_2) \cdot \widehat{\mathbf{e}}^0_1 &&= T^a_{12} \\
    \frac{\partial u_2}{\partial x_2} &= (\mathbf{e}_2 - \mathbf{e}^0_2) \cdot \widehat{\mathbf{e}}^0_2 &&= T^a_{22} - 1
\end{align*}
$\epsilon$ can then be retrieved from $T^a$.

In practice, we calculate basis lattice vectors of the reciprocal lattice $g = (\mathbf{g}_1,\mathbf{g}_2)$, by performing an intensity-weighted fit to measured Bragg peak positions at each scan position.
The corresponding reference vectors $g^0 = (\mathbf{g}^0_1,\mathbf{g}^0_2)$ can be determined several ways, including defining a reference region, dataset, or using a known crystal structure.
At this point it is possible to use $g$ to determine $a$, then compute $\epsilon$ with the methods above.
Alternatively, assuming sufficiently small deformations that we may discard terms of second order and higher in a Taylor expansion in both rotation and scaling, it is possible to compute $\epsilon$ directly from the transformation $T_{g^0 \to g}$, describing the linear deformation of reciprocal space, with remarkably little alteration to the above equations.
In this case, the final expressions for the strain are
\begin{eqnarray}
\label{E:final_crystal_strain}
    \epsilon_{11} &=& 1 - T^g_{11} \nonumber \\
    \epsilon_{22} &=& 1 - T^g_{22} \nonumber \\
    \epsilon_{12} &=& - \frac{1}{2}\left(T^g_{12} + T^g_{21}\right) \nonumber \\
    \theta_{R}    &=& \frac{1}{2}\left(T^g_{12} - T^g_{21}\right)
\end{eqnarray}

To measure strain from the diffraction pattern of an amorphous material, we fit ellipses at each probe position using the methods described in Appendix~\ref{A:ellipses}. After shifting the origin of each ellipse to $\mathbf{k} = (0,0)$, we have 
\begin{equation}
\label{E:amor_strain_01}
    {q_{\rm{ref}}}^2 = 
    A {k_x}^2 + B k_x k_y + C {k_y}^2 
\end{equation}
where $q_{\rm{ref}}$ is a reference radius, which defines an undeformed (circular) amorphous halo given by
\begin{equation}
\label{E:amor_strain_02}
    {q_{\rm{ref}}}^2 = 
    {k_x}^2 + {k_y}^2. 
\end{equation}

Because the measurement takes place in reciprocal space, it is more convenient to define the transformation matrix from the measured ellipse given by Eq.~\ref{E:amor_strain_02} to the reference circle given by Eq.~\ref{E:amor_strain_01}, which is given by
\begin{gather}
    \begin{bmatrix} 
        {k_x}' \\  {k_y}'
    \end{bmatrix}
    = \mathbf{T}
    \begin{bmatrix}
        k_x \\ k_y
    \end{bmatrix},
\end{gather}
where
\begin{gather}
    \mathbf{T} =
    \frac{1}{\sqrt{A + C + W}}
    \begin{bmatrix}
    A + \frac{1}{2} W & 
    B \\
    B & 
    C + \frac{1}{2} W \\
   \end{bmatrix},
   \nonumber
\end{gather}
where
\begin{equation}
    W = \sqrt{4 A C - B^2}.
   \nonumber
\end{equation}
This expression is valid as long as the roots are real, i.e. $4 A C - B^2 > 0$. To calculate the strain deformation tensor, we proceed in a similar manner to \ref{E:final_crystal_strain}, althrough we note the direction of the transformation has already been changed to the real space transformation directions,
\begin{eqnarray}
    \epsilon_{11} &=& T_{11} - 1
    \nonumber \\
    \epsilon_{22} &=& T_{22} - 1
    \nonumber \\
    \epsilon_{12} &=& \frac{1}{2}\left(T_{12} + T_{21}\right).
\end{eqnarray}
The full expressions for the strain tensor components are
\begin{eqnarray}
    \epsilon_{11} &=&
    \frac{A + \frac{1}{2}W}{\sqrt{A + C + W}} - 1
    \nonumber \\
    \epsilon_{22} &=& 
    \frac{C + \frac{1}{2}W}{\sqrt{A + C + W}} - 1
    \nonumber \\
    \epsilon_{12} &=& 
    \frac{B}{\sqrt{A + C + W}}
\end{eqnarray}
Taking a first order Taylor expansion about $A=1$, $C=1$, and $B=0$ yields the linear strain approximation
\begin{eqnarray}
    \epsilon_{11} &=& \frac{1}{2}(A - 1)
    \nonumber \\
    \epsilon_{22} &=& \frac{1}{2}(C - 1 )
    \nonumber \\
    \epsilon_{12} &=&  \frac{1}{2} B.
\end{eqnarray}
Note that when using the linear approximation above, it is important to use a value for $q_{\rm{ref}}$ that is very close to the reference lattice average scattering radius, as the accuracy of the above expressions will suffer as the approximations $A \approx 1$ and $C \approx 1$ become worse. 

An alternative method of determining the real-space strains corresponding to an ellipse can be done using matrix notation. In matrix form, the ellipse equation~\ref{E:amor_strain_01} can be represented as 
\begin{equation}
    \label{E:strain_matrix2}
    M = \begin{bmatrix} A & B/2 \\ B/2 & C \end{bmatrix},
\end{equation}
where the major and minor axis directions are the eigenvectors of $M$, and their lengths are the square root of the eigenvalues. In the eigenbasis reference frame then, the transformation matrix, $\mathbf{T}$, is simply the square root of the diagonalized eigenvalues, aligned with the corresponding eigenvectors. However, this must be rotated back to the traditional $xy$ reference frame, or another chosen reference frame, via tensor rotation
\begin{equation}
    \mathbf{T}' = \mathbf{R}\mathbf{T}\mathbf{R}^{\rm{T}}
\end{equation}
where $\mathbf{R}$ is a standard rotation matrix and the superscript T represents transpose. The angle with respect to the $xy$ axis can be found by taking the two-argument arctangent ($\mathrm{atan}2$) of an eigenvector. Finally, once $\mathbf{T}'$ is in the correct orientation, the strains are simply
\begin{eqnarray}
    \epsilon_{11} &=& T_{11} - 1
    \nonumber \\
    \epsilon_{22} &=& T_{22} - 1
    \nonumber \\
    \epsilon_{12} &=& \left(T_{12} + T_{21}\right).
\end{eqnarray}

We also note that most experiments will contain some degree of ellipticity even when no strain is present. In these cases, we will subtract the reference strain state from all measurements.


\section{Radial Distribution Functions}
\label{A:RDF}

We computation the radial distribution function following the methods of \cite{mu2016radial, mitchell2012rdftools}.
Beginning from an amorphous diffraction pattern, often averaged over many probe positions to increase the signal to noise ratio, we measure  the radial intensity $\langle I(k) \rangle_\phi$ averaged over the angular direction $\phi$.
Figure~\ref{F:RDF}a-c shows such data for amorphous silicon. 
Next, we estimate the structure factor using the expression
\begin{equation}
\label{E:rdf_structure_factor}
    \Phi(q) = \frac{
    \langle I(q) \rangle_\phi
    - I_{\rm{BG}}(q)  
    - N \langle {f(q)}^2 \rangle}{
     N \langle {f(q)}^2 \rangle} q \, M(q),
\end{equation}
where $I_{\rm{BG}}(q)$ is a background intensity estimate, $\langle {f(q)}^2 \rangle$ is the mean-square of the parameterized single-atom scattering factors for all atomic species present, multiplied by $N$ total atoms in the probe volume, and $M(q)$ is a masking function.
The single atom scattering term is typically fit to the high scattering angle region, past the region where oscillating structure factor peaks are visible.
The background $I_{\rm{BG}}(q)$ can be an additional constant offset, a more complex fitting function, or just neglected.
A masking envelope function $M(q)$ is required to zero the structure factor $\Phi(q)$ at low scattering angles due to residual intensity from the central beam.
This function can also be used to zero the structure factor $\Phi(q)$ at high scattering angles as well, due to residual fitting errors in $N \langle {f(q)}^2 \rangle$ and $ I_{\rm{BG}}(q)$.
Figure~\ref{F:RDF}c shows the single atom scattering fit, and Fig.~\ref{F:RDF}d shows the masked structure factor.

Next, we obtain the reduced radial distribution function (RDF) $g(r)$ by taking the discrete (type II) sine transform of $\Phi(q)$, equal to
\begin{equation}
    g(r) = \sum_{q=0}^{q_{\rm{max}}}
    \Phi(q)  \sin \left[
    \frac{\pi}{q_{\rm{max}}}
    \left( \frac{q}{\Delta q} + \frac{1}{2} \right)
    \left( \frac{r}{\Delta r} + 1 \right)
    \right],
\end{equation}
where $q_{\rm{max}}$ is the maximum $q$ where $\Phi(q)$ is non-zero, and $\Delta q$ and $\Delta r$ are the pixel sizes in diffraction and real space respectively. 
$g(r)$ should ideally approach 0 as $r \rightarrow 0$ within the nearest neighbor shell, and approach 1 as $r \rightarrow \infty$, but errors in the above fitting procedure can cause deviations from these results. \cite{mu2016radial} 
Other authors recommend subtracting a $4^{\rm{th}}$ order polynomial from $\Phi(q)$ in order to reduce low spatial frequency artifacts. 
Here, we add a sigmoidal damping mask which clamps the RDF to zero as $r \to 0$.
Figure~\ref{F:RDF}e shows the final result of the calculation, $g(r)$.

Finally, we can also compute the atomic density $\rho(r)$ of our sample using the expression
\begin{equation}
    \rho(r) = r \left[
    g(r) + 4 \pi r \rho_0
    \right],
\end{equation}
where $\rho_0$ is the bulk atomic density of the sample. This expression can be used to determine the coordination number of neighboring shells of atoms, by integrating over the distances $r$ corresponding to a specific shell.

\section{Fluctuation Electron Microscopy}
\label{A:FEM}

The FEM computation here follows the methods of \cite{bogle2010size}.
The first steps are identical to an RDF study, namely calibrating the elliptic distortions, performing the polar transformation and measuring the average intensity as a function of scattering angle $\langle I(q) \rangle_{\phi,\mathbf{R}}$ from all diffraction patterns at probe positions $\mathbf{R}$.
Next, we measure the variance $V(q)$ of the intensity as a function of scattering angle over all diffraction patterns
\begin{equation}
    V(q) = 
    \langle
    \left[
    \langle I(q) \rangle_{\phi}  -
    \langle I(q) \rangle_{\phi,\mathbf{R}} 
    \right]^2
    \rangle_{\mathbf{R}}.
\end{equation}
In order to reduce the effect of thickness when comparing multiple datasets, we then compute the normalized variance $V_{\rm{norm}}(q)$ as
\begin{equation}
    V_{\rm{norm}}(q) = 
    \frac{
    V(q)
    }{
    \langle I(q) \rangle_{\phi,\mathbf{R}}^2
    }
\end{equation}
The RDF measurement described above is primarily sensitive to the 2-body atomic pair correlations, whereas the FEM variance is more sensitive to 4-body pair-pair correlations \cite{treacy1996variable}.

\section{Differential Phase Contrast}

\label{A:dpc}

In this appendix we summarize the mathematical underpinnings of the differential phase contrast method, adapted from Ref.~\cite{waddell1979linear}.

Following from Appendix~\ref{A:STEM}, we select for the STEM detector in Eq.~\ref{E:stem_imaging}, the so-called first moment detector function $\mathbf{D}(k) = \mathbf{k}$.
Note that, in contrast to Eq.~\ref{E:stem_imaging}, here we use a vector valued detector function, yielding a vector valued image:
\begin{equation}
\label{E:dpc_detector}
    \mathbf{I}(\mathbf{R}) = \int \lvert \Psi(\mathbf{k},\mathbf{R}) \rvert^2 \mathbf{k} d\mathbf{k}
\end{equation}
Physically, this distinction reflects that such a detector is sensitive to deflections of the total intensity of the electron beam, which is itself a vector quantity.
A good approximation to Eq.~\ref{E:dpc_detector} is possible using a segmented detector geometries~\cite{muller2019comparison}.

Substituting this into Eq.~\ref{E:diffraction}, this detector choice allows us to write:
\begin{align*}
    \mathbf{I}(\mathbf{R}) =& \int \lvert \mathcal{F}_{\mathbf{r}\to\mathbf{k}}\left( \psi(\mathbf{r},\mathbf{R}) \right) \rvert^2 \mathbf{k} d\mathbf{k} \\
    =& \int
    \mathcal{F}_{\mathbf{r}\to\mathbf{k}}\left(\psi(\mathbf{r},\mathbf{R}) \right)
    \mathcal{F}^*_{\mathbf{r}\to\mathbf{k}}\left(\psi(\mathbf{r},\mathbf{R}) \right) \mathbf{k} d\mathbf{k} \\
    =& \frac{1}{2\pi i} \int
    \mathcal{F}_{\mathbf{r}\to\mathbf{k}}\left(\mathbf{\nabla}_\mathbf{r}\psi(\mathbf{r},\mathbf{R}) \right) \\
    &\times \mathcal{F}^*_{\mathbf{r}\to\mathbf{k}}\left(\psi(\mathbf{r},\mathbf{R}) \right) d\mathbf{k}
\end{align*}
where $^*$ indicates a complex conjugation, and in the last line the derivative property of the Fourier transform has been invoked.\footnote{I.e. $\mathcal{F}_{\mathbf{r}\to\mathbf{k}}(\mathbf{\nabla}f(\mathbf{r})) = 2\pi i\mathbf{k}\mathcal{F}_{\mathbf{r}\to\mathbf{k}}(f(\mathbf{r}))$}

So far no assumptions have been made.
If a thin specimen is assumed\footnote{A necessary but not sufficient condition is that the probe depth of field (equal to $1.7\lambda/\alpha^2$) is much greater than the specimen thickness.} we may model the probe-sample interaction as multiplication of the electron wave function with a specimen transmission function $T(\mathbf{r})$ and we can continue with
\begin{align}
    \mathbf{I}(\mathbf{R}) =& \frac{1}{2\pi i} \int d\mathbf{k} \int d\mathbf{r} 
    e^{-2\pi i\mathbf{k}\cdot\mathbf{r}} \nonumber \\ 
    & \biggl(\psi_0(\mathbf{r}-\mathbf{R})\mathbf{\nabla}T(\mathbf{r})
    + T(\mathbf{r})\mathbf{\nabla}\psi_0(\mathbf{r}-\mathbf{R}) \biggr) \nonumber\\
    & \times \int d\mathbf{r}' e^{2\pi i\mathbf{k}\cdot\mathbf{r}'} \psi^*_0(\mathbf{r}-\mathbf{R})T^*(\mathbf{r}') \nonumber \\
    =& \frac{1}{2\pi i} \int d\mathbf{r}
    \biggl(\lvert\psi_0(\mathbf{r}-\mathbf{R})\rvert^2 \mathbf{\nabla}T(\mathbf{r})T^*(\mathbf{r}) \nonumber\\
    & + \mathbf{\nabla}_\mathbf{r}\psi_0(\mathbf{r}-\mathbf{R})\psi^*_0(\mathbf{r}-\mathbf{R})\lvert T(\mathbf{r})\rvert^2 \biggr) \label{E:DPC_intermediatestep1}
\end{align}

We now take the phase object approximation and write the transmission function as $T(\mathbf{r}) = e^{i\sigma V(\mathbf{R})}$.
The second half of the sum in Eq.~\ref{E:DPC_intermediatestep1} then becomes 
\begin{align*}
    \frac{1}{2\pi i} & \int d\mathbf{r} \mathbf{\nabla}_\mathbf{r}\psi_0(\mathbf{r}-\mathbf{R})\psi^*_0(\mathbf{r}-\mathbf{R}) \\
    =& \frac{\hbar}{2\pi} \int d\mathbf{r} \psi^*_0(\mathbf{r}-\mathbf{R}) \mathbf{\widehat{p}} \psi_0(\mathbf{r}-\mathbf{R}) \\
    =& \frac{\hbar}{2\pi} \langle \mathbf{p} \rangle
\end{align*}
where $\langle\mathbf{p}\rangle$ is the expectation value of the probe momentum, and we've made use of the fact that the momentum operator is $\mathbf{\widehat{p}} = \frac{1}{i\hbar}\mathbf{\nabla}$.
But $\langle\mathbf{p}\rangle$ is independent of $\mathbf{R}$ and thus provides some constant offset to $\mathbf{I}(\mathbf{R})$, and can be neglected.

Continuing from Eq.~\ref{E:DPC_intermediatestep1} and again using the phase object approximation, we find
\begin{align}
    \mathbf{I}(\mathbf{R}) =& \frac{1}{2\pi i} \int d\mathbf{r}
    \lvert\psi_0(\mathbf{r}-\mathbf{R})\rvert^2
    \left(i\sigma \mathbf{\nabla}V(\mathbf{r})\right) \nonumber \\
    =& -\frac{\sigma}{2\pi} \int d\mathbf{r} \mathbf{E}(\mathbf{r}) 
    \lvert\psi_0(\mathbf{r}-\mathbf{R})\rvert^2 \nonumber \\
    \label{eq:EconvProbe}
    =& -\frac{\sigma}{2\pi} \mathbf{E}(\mathbf{R}) * \lvert\psi_0(\mathbf{R})\rvert^2
\end{align}
where $*$ denotes a convolution, and we've made use of the fact that $\mathbf{E} = -\mathbf{\nabla}V$.

\begin{figure}[t]
    \centering
    \includegraphics[width=\columnwidth]{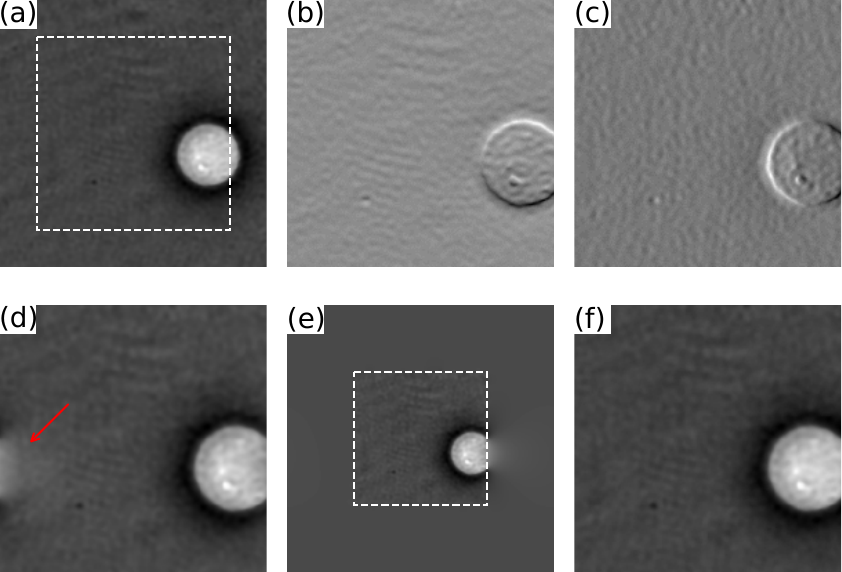}
    \caption{(a) A test image, a holographic reconstruction of a biological cell in saline from Ref.~\cite{Muller2018Phase}, along with the numerical derivatives of a cropped region (white dashed line) of the image in (b) and (c). A DPC reconstruction from (b) and (c) produces (d), which has artifacts, most noticeably that indicated with a red arrow, result form the boundary conditions implicitly assumed by the use of fast Fourier transforms. (e) The approach implemented by py4DSTEM, discussed in the body text, is to reconstruct on a padded grid [the original grid of (b) and (c) is indicated by the dashed outline]. The result of 10 iterations is shown in (f) which does not exhibit the same artefacts as (d). }
    \label{fig:DPCperiodic}
\end{figure}

To solve for $V$ given $I$ requires that we integrate Eq~(\ref{eq:EconvProbe}), and we follow \cite{arnison2004linear,close2015towards,lazic2016phase} and write
\begin{align}
    \label{eq:FFTsolutionDPC}
    V(\mathbf{R}) = - \mathcal{F}^{-1}_\mathbf{k\rightarrow r} \left[\frac{\mathbf{k}\cdot\mathcal{F}_\mathbf{k\rightarrow r}\{\mathbf{I}(\mathbf{R})\}}{ i k^2}\right]/\sigma\,.
\end{align}
Technically a quantitative measurement of $V(\mathbf{R})$ also requires deconvolution of the probe wave function~\cite{brown2017measuring}, a step which we ignore in py4DSTEM due to the fact that this serves to amplify noise in most experimental datasets.

An additional challenge is the implicit assumption of periodic boundary conditions in the use of fast Fourier transforms to solve Eq.~(\ref{eq:FFTsolutionDPC}). This is demonstrated in Fig.~\ref{fig:DPCperiodic} for reconstruction of a test image, a holographic reconstruction of a  biological cell in saline from Ref.~\cite{Muller2018Phase}, shown in Fig.~\ref{fig:DPCperiodic}(a).
The $x$ and $y$ numerical derivatives of a cropped region (indicated by a white dashed outline) of Fig.~\ref{fig:DPCperiodic}(a) are shown in Fig.~\ref{fig:DPCperiodic}(b) and (c) respectively and the result of the solution proposed in Eq.~(\ref{eq:FFTsolutionDPC}) is shown in Fig.~\ref{fig:DPCperiodic}(d).
An artefact of the implicit periodic boundary conditions can be seen in part of the white cell on the right-hand side of the frame appearing on the opposite left-hand side as indicated by a red arrow in Fig.~\ref{fig:DPCperiodic}(d).
The approach pursued in py4DSTEM to remedy this is to solve Eq.~(\ref{eq:FFTsolutionDPC}) on a larger grid, one that has been ``padded'' with zeros as shown in Fig.~\ref{fig:DPCperiodic}(e). 
The original grid corresponding to the input derivatives in Fig.~\ref{fig:DPCperiodic} is indicated by a white dashed outline in Fig.~\ref{fig:DPCperiodic}.
The gradient of the solution in Fig.~\ref{fig:DPCperiodic}(e) is taken and the result subtracted from the input derivatives Fig.~\ref{fig:DPCperiodic}(b) and (c).
The result of this subtraction (the residual) forms the input for another solution of Eq.~(\ref{eq:FFTsolutionDPC}) on a padded grid which is added to the phase solution as a correction.
This processes is iterated upon until convergence, the result of just 10 iterations is shown in Fig.~\ref{fig:DPCperiodic}(f), a more faithful reproduction of Fig.~\ref{fig:DPCperiodic}(a) than Fig.~\ref{fig:DPCperiodic}(d).

\section{Ptychography}
\label{A:ptychography}

In this appendix we summarize the mathematics of the ptychographic reconstruction method implemented in Fig.~\ref{F:dpc}.
Beginning from the transmission function approximation, the measured intensity $I$ at a spatial frequency $\mathbf{k}$ and a probe position $\mathbf{R}$ is given by
\begin{equation}
    I(\mathbf{k},\mathbf{R}) = \left|\int \psi_0(\mathbf{r}-\mathbf{R})T(\mathbf{r}) e^{2 \pi i \mathbf{r}\cdot\mathbf{k}} d\mathbf{r}\right|^2.
\end{equation}
The goal is to retrieve $T$ from $I$.

We first consider the transformed datacube $I(\mathbf{k},\mathbf{K}) = \mathcal{F}_{\mathbf{R}\to\mathbf{K}}I(\mathbf{k},\mathbf{R})$, where $\mathbf{K}$ is the reciprocal coordinate to scan position $\mathbf{R}$.
Thus this is the datacube has been written in terms of scan frequencies.
Assuming the sample is a weak phase object, this may be written as \cite{rodenburg1993experimental}
\begin{align}
\label{E:g_function}
    I(\vect{k},\vect{K}) =& \left|\Psi_{0}(\vect{k})\right|^2\delta(\mathbf{k}) \nonumber\\ 
    &+ \Psi_{0}(\vect{k})\Psi_{0}^*(\vect{k}+\vect{K})T(-\mathbf{k})^* \nonumber\\
    &-\Psi_{0}^*(\mathbf{k})\Psi_{0}(\mathbf{k}-\mathbf{K})T(\mathbf{k})
\end{align}
where $T(\mathbf{k}) = \mathcal{F}_{\mathbf{r}\to\mathbf{k}}T(\mathbf{r})$.
The latter two terms in this expression each contain two copies of the aperture function, one centered at the optic axis and one shifted by the scan frequency $\mathbf{K}$.
For some given $\mathbf{K}$, these terms can each be nonzero only at values of $\mathbf{k}$ where both disks are nonzero; that is, in the overlap between the shifted and unshifted disks.
By looking only at the nonzero overlap between two disks, it is possible to simplify and solve Eq.~\ref{E:g_function} by eliminating one of its terms.

To eliminate the third term, define the set of pixels
\begin{align*}
    \mathcal{K} = \{\mathbf{k}\,:\,&(\lvert\mathbf{k}\rvert < k_0) \\  &\land\,(\lvert\mathbf{k}+\mathbf{K}\rvert < k_0) \\
    &\land\,(\lvert\mathbf{k}-\mathbf{K}\rvert > k_0)\}
\end{align*}
where $\land$ is the logical and operation and the maximum disk size is $k_0 = \frac{\alpha}{\lambda}$ for convergence semi-angle $\alpha$ and electron wavelength $\lambda$.
Here, the first line requires that $\mathbf{k}$ is inside the central disk, the second line requires that $\mathbf{k}$ is inside the disk shifted by $-\mathbf{K}$, and the third line requires that $k$ is also \textit{outside} the disk shifted by $\mathbf{K}$.
Thus $\mathcal{K}$ selects a region of double overlap while also excluding the region of triple overlap.

If only data from $\mathbf{k} \in \mathcal{K}$ is used, the third term in Eq.~\ref{E:g_function} vanishes, so that 
\begin{equation}
    T(-\mathbf{k})^* = \sum_{\mathbf{k} \in \mathcal{K}} \frac{I(\mathbf{k},\mathbf{K})}{\Psi_{0}(\vect{k})\Psi_{0}^*(\mathbf{k}+\mathbf{K})},
\end{equation}
and $T(\mathbf{r})$ can be obtained with a subsequent inverse Fourier transform.
Since this uses only data from a single double-overlap region, this method has been dubbed single-sideband reconstruction.

This approach can be extended to include data from the whole bright-field in the reconstructions.
If the sample is a weak phase object, it obeys Friedel symmetry, so that $T(\mathbf{k}) = T(-\mathbf{k})^*$ \cite{rodenburg1993experimental}.
Inserting this into Eq.~\ref{E:g_function} and solving gives \cite{yang2016enhanced}
\begin{equation}
    T(\mathbf{k}) = \sum_{\mathbf{k}\,:\,\lvert\mathbf{k}\rvert < k_0}I(\mathbf{k},\mathbf{K})\cdot\frac{\Gamma^*(\mathbf{k},\mathbf{K})}{\left|\Gamma(\mathbf{k},\mathbf{K})\right|^2},
\end{equation}
where $\Gamma$ is the disk-overlap function
\begin{equation}
    \label{E:diskoverlap}
    \Gamma(\mathbf{k},\mathbf{K}) = \Psi_{0}(\mathbf{k})\Psi_{0}^*(\mathbf{k}+\mathbf{K}) - \Psi_{0}^*(\mathbf{k})\Psi_{0}(\mathbf{k}-\mathbf{K}).
\end{equation}




\bibliography{4DSTEMrefs}
\end{document}